\documentclass[journal,11pt,draftclsnofoot,onecolumn]{IEEEtran}
\usepackage{graphicx}
\usepackage[square,comma,numbers,sort&compress]{natbib}
\usepackage{amsthm,amssymb,amscd,amssymb,amsmath,mathrsfs}
\usepackage{epstopdf}
\usepackage{balance}
\usepackage{multirow} %multirow for format of table
\usepackage{amsmath}
\usepackage{xcolor}
\usepackage{subfigure}
\usepackage{enumerate}

\makeatletter
\newcommand{\rmnum}[1]{\romannumeral #1}
\newcommand{\Rmnum}[1]{\expandafter\@slowromancap\romannumeral #1@}
\makeatother

\theoremstyle{plain}
\newtheorem{thm}{Theorem}

\newtheorem{lem}{Lemma}

\theoremstyle{definition}

\theoremstyle{remark}
\newtheorem{remark}{Remark}

\begin{document}
\title{Supervisor Localization of Discrete-Event Systems with Infinite Behavior\\
 \Large(\today)
 }

\author{Renyuan Zhang$^{1}$, Kai Cai$^{2}$% <-this % stops a space
%\thanks{*This work was supported in part by the National Nature Science Foundation of China, Grant no. 61403308;
%the Natural Science Foundation of Shaanxi Province, China, Grant no. 2017JM5061;
%JSPS KAKENHI Grant no. JP16K18122.
%}% <-this % stops a space
\thanks{$^{1}$R. Zhang is with School of Automation, Northwestern Polytechnical University, China. Email:
        {\tt\small ryzhang@nwpu.edu.cn}.}%
\thanks{$^{2}$K. Cai is with Department of Electrical and Information Engineering, Osaka City University, Japan. Email:
        {\tt\small kai.cai@eng.osaka-cu.ac.jp}.}%
}

\maketitle

\begin{abstract}
Recently we developed \emph{supervisor localization}, a top-down approach to
distributed control of discrete-event systems (DES) with \emph{finite} behavior.
Its essence is the allocation of monolithic (global) control action
among the local control strategies of individual agents. In this report,
we extend supervisor localization to study the distributed control of
DES with \emph{infinite} behavior. Specifically, we first employ Thistle and Wonham's
supervisory control theory for DES with infinite behavior to compute
a \emph{safety} supervisor (for safety specifications) and a \emph{liveness} supervisor
(for liveness specifications), and then design a suitable localization
procedure to decompose the safety supervisor into a set of safety local controllers, one for
each controllable event, and decompose the liveness supervisor into a set of
liveness local controllers, two for each controllable event. The localization procedure
for decomposing the liveness supervisor is novel; in particular, a local controller
is responsible for disabling the corresponding controllable event on only part of
the states of the liveness supervisor, and consequently, the derived local controller
in general has states number no more than that computed by considering the disablement on
all the states. Moreover, we prove that the derived local controllers achieve the same
controlled behavior with the safety and liveness supervisors. We finally
illustrate the result by a Small Factory example.
\end{abstract}


\section{Introduction}

%\section{Introduction}

In \cite{CaiWon10a,CaiWon10b,ZhangEt13,CaiWon15,CaiWon16,ZhangCW17}
we developed a top-down approach, called {\it supervisor localization}, to the
distributed control of multi-agent discrete-event systems (DES). % with {\it finite} behaviors.
This approach first synthesizes a monolithic supervisor (or a
heterarchical array of modular supervisors), and then decomposes the supervisor into a set of
local controllers for the component agents. Localization
creates a purely distributed control architecture in which each
agent is controlled by its own local controller; this is particularly
suitable for applications consisting of many autonomous components,
e.g. multi-robot systems. Moreover, localization can significantly
improve the comprehensibility of control logic, because the resulting local
controllers typically have many fewer states than their parent supervisor.

These works focus on DES with {\it finite} behaviors \cite{Wonham16a}, in which DES are modelled as generators accepting
$*$-languages (consisting of finite-length strings) and the specifications are
expressed by $*$-languages. In modelling and control of reactive systems
(e.g. automated factories, operating systems, communication protocols), however,
the systems may operate indefinitely, and the specifications may require that
every system component must operate infinitely often.
In these cases, {\it $\omega$-automata} on infinite inputs and {\it $\omega$-languages}
consisting of infinite-length strings were introduced to model the DES with
{\it infinite} behavior and specify the specifications respectively. Notable
works on synthesizing supervisors for the DES with infinite behavior include the following. 
First, Ramadge \cite{Ramadge89} models the
DES with infinite behavior by B\"uchi automata and derives conditions ($*$-controllability
and $\omega$-closure) for the existences of supervisors; within the same framework,
Young et al. \cite{YoungEt92} derives another supervisor existence condition (replacing
$\omega$-closure by finite stabilizability) under less restrictive conditions.
Then, Thistle and Wonham \cite{Thistl91,ThiWon94a,ThiWon94b} introduce the concept
of $\omega$-controllability which is closed under arbitrary set union, and develop
a procedure to synthesize supervisors satisfying given specifications
expressed by $\omega$-languages; Kumar et al. \cite{Kumar1992} proposed an
alternative algorithm to compute the supremal $\omega$-controllable sublanguage.
Later, Thistle \cite{Thistle1995} extend the result in \cite{Thistl91} to a more 
general case where the plant DES are
modelled by deterministic Rabin-automata. More recently, Thistle and Lamouchi
\cite{Thistle2009} addressed the issue of partial observation in
the supervisory control of DES with infinite behavior. To the best of our knowledge, however,
there is no result on distributed control for multi-agent DES with infinite behavior
reported in the literature.

In this paper, we extend supervisor localization to address distributed
control for DES with infinite behavior. Our approach is as follows.
Given a DES plant with infinite behavior and {\it safety} and {\it liveness}
specifications, we first synthesize a safety supervisor (for safety specifications)
and a liveness supervisor (for liveness specifications) by the method proposed
by Thistle and Wonham \cite{Thistl91,ThiWon94b}.
The infinite controlled behavior of the plant is restricted through the
control actions on finite strings, thus as in DES with finite behavior \cite{Wonham16a}, 
we implement the supervisors by $*$-automata.
We then adopt the localization procedure in \cite{CaiWon10a} with suitable modifications to decompose the
automata-based safety and liveness supervisors into local controllers for individual controllable events.
Moreover we prove that the derived local controllers are control equivalent to the
synthesized safety and liveness supervisors.

The contributions of this paper are twofold.
First, we develop a new supervisor localization theory for DES with infinite behavior 
in Thistle and Wonham's supervisory control framework \cite{ThiWon94b}, which
supplies a systematic, computationally effective approach to distributed
control of multi-agent DES with infinite behavior.
In particular, we first decompose the safety supervisor into a set of local controllers,
one for each controllable event, by the localization procedure in \cite{CaiWon10a};
then we decompose the liveness supervisor into a set of local controllers, however,
two for each controllable events, by a newly developed localization procedure.
The central idea of the new procedure is the
new definition of disabling function with a new language: only the disablement on part of the states are
defined, i.e. an event is defined as disabled at one state only if the state can be visited
by strings in the given language. With this new disabling
function, we define new concepts of control consistency and control cover, and the resultant
local controllers in general have states number no more than that computed by the 
localization procedure in \cite{CaiWon10a} where the disablement on all the states
are considered.

Second, we identify the essence of localization procedure for DES with infinite behavior:
only the disabling/enabling actions on finite strings need be considered.
Namely, if the control equivalence of the local controllers
with their parent supervisors on finite behavior is guaranteed, the control equivalence
on infinite behavior can be derived by Lemma~\ref{lem:equ_lim} in Section~\ref{Subsec:MainResult}, which
declares that the operator limit (mapping finite strings to infinite
strings whose prefixes are all contained in the given finite strings) will not change
the language equivalence on intersections. Consequently, control
consistency relation and control cover, the central concepts of the localization procedure,
are defined only on the disabling and enabling functions, irrelevant to the infinite behaviors.
We demonstrate the above result by a case study of Small Factory example \cite{Thistl91}.

Our proposed localization procedure can in principle be used to construct
local controllers from supervisors computed by any other synthesis
method for DES with infinite behavior e.g. \cite{Ramadge89,YoungEt92,Thistle1995}.
In this paper, we adopt the Thistle and Wonham's supervisory control theory for
two reasons. First, it extends basic results of the supervisory control theory of Ramadge
and Wonham \cite{RamWon87,Wonham16a} for DES with finite behavior to infinite behavior,
and generalizes results of \cite{Ramadge89} to the case in which specification languages
need not be $\omega$-closed relative to plant behavior. Second, the supervisors
synthesized by Thistle and Wonham's theory can be implemented by $*$-automata,
which are eligible to be decomposed into local controllers by our previous work on supervisor
localization procedure with appropriate modifications.

The paper is organized as follows. Section \ref{Sec:prelin} reviews the preliminaries
on DES with infinite behavior and Thistle and Wonham's supervisory control theory.
Section \ref{Sec:ProbForm} formulates the problem of Supervisor Localization for DES with infinite behavior.
Section \ref{Sec:suploc} presents the localization procedure and proves the control equivalence
of the derived local controllers with their parent supervisors, and Section
\ref{Sec:CaseStudy} illustrates the proposed localization procedure by a Small Factory example.
Finally Section \ref{Sec:Concl} states our conclusions.

%% the end %%

\section{Preliminaries on DES with Infinite Behavior} \label{Sec:prelin}

In this section, we briefly review Thistle and Wonham's supervisory
control framework of discrete-event systems (DES) with infinite
behavior \cite{ThiWon94a,ThiWon94b,Thistl91}.

% the end

\subsection{Discrete-Event Systems with Infinite Behavior} \label{Subsec:DESModel}

A discrete-event system (DES) with infinite behavior (plant to be controlled)
is modeled as a deterministic B\"{u}chi automaton\footnote{The DES with infinite
behavior can also be modeled by other form of $\omega$-automata with different types
of acceptance criteria, e.g. Muller automata, Rabin automata, Street automata.
It is known \cite{Mukund12} that deterministic B\"{u}chi automata represent a strict
subset of $\omega$-regular languages, having less expressive power
than nondeterministic B\"{u}chi automata, deterministic and nondeterministic
Muller automata, and deterministic and nondeterministic Rabin automata which
represent the full set of $\omega$-regular languages. In this report, following
Thistle and Wonham's framework \cite{Thistle1994b}, we focus on
the subset of $\omega$-regular languages that are represented by deterministic B\"{u}chi
automata, and leave the extension to the full set for future work.}
\begin{align} \label{eq:plant}
{\bf G}: = (Q, \Sigma, \delta, q_0, \mathcal{B}_{Q}),
\end{align}
where $Q$ is the finite state set, $q_0$ is the initial state, $\Sigma$ is the
finite event set (alphabet), $\delta: Q \times \Sigma \rightarrow Q$ is the (partial)
state transition function, and $\mathcal{B}_{Q} \subseteq Q$
is the B\"uchi acceptance criterion. In the usual way, $\delta$ is extended
to $\delta:Q\times\Sigma^* \rightarrow Q$, and we write $\delta(q,s)!$ to mean
that $\delta(q,s)$ is defined. Let $\Sigma^*$ be the set of all finite strings over $\Sigma$,
including the empty string $\epsilon$, and $\Sigma^\omega$ the set of
all infinite strings over $\Sigma$; the disjoint union of $\Sigma^*$ and $\Sigma^\omega$ is denoted by
$\Sigma^\infty$, i.e. $\Sigma^\infty = \Sigma^* \dot\cup \Sigma^\omega$.
The DES ${\bf G}$ has both finite behavior and infinite behavior. The {\it finite behavior} of
${\bf G}$ is the {\it $*$-language} $L({\bf G}) \subseteq \Sigma^*$ accepted by the $*$-automaton
$(Q, \Sigma, \delta, q_0)$, i.e. \[L({\bf G}) := \{s \in \Sigma^* | \delta(q,s)! ~\& ~ \delta(q,s) \in Q\};\]
and the {\it infinite behavior} of ${\bf G}$ is the $\omega$-language $S({\bf G})$ accepted by
the $\omega$-automaton ${\bf G}$ with B\"{u}chi acceptance criterion $\mathcal{B}_{Q}$, i.e.
\[S({\bf G}) : = \{s \in \Sigma^\omega| \Omega(s) \cap \mathcal{B}_{Q} \neq \emptyset\}\]
where $\Omega(s)$ is set of states that $s$ visits infinitely often. %We also write
%${\mathcal{A}}$ as $(L, S)$, where $L$ and $S$ denote the $*$-behavior and $\omega$-behavior of ${\mathcal{A}}$, respectively;
%namely, $\mathcal{A} = (L, S)$, where $L := L^*({\mathcal{A}})$ and $S := L^\omega({\mathcal{A}})$.

A string $s \in \Sigma^*$ is a {\it prefix} of a string $v \in \Sigma^\infty$, written
$s \leq v$, if there exists $t \in \Sigma^\infty$ such that $v = st$.
The {\it (prefix) $*$-closure} of a language $K \subseteq \Sigma^\infty$ is defined by
\begin{align}\label{eq:prefix}
pre(K) := \{s \in \Sigma^* | (\exists s_1 \in K) ~s \leq s_1\}
\end{align}
If $K = pre(K)$, we say that $K$ is $*$-closed. In this report, we assume that (i) $pre(L({\bf G})) = L({\bf G})$,
i.e. $L(\bf G)$ is $*$-closed, and (ii) $pre(S({\bf G})) = L({\bf G})$, i.e.
${\bf G}$ is {\it deadlock-free}. Define the {\it limit} of a $*$-language $K$ by
\begin{align} \label{eq:def_lim}
lim(K) := pre^{-1}(K) \cap \Sigma^\omega
\end{align}
where $pre^{-1}(K) := \{v\in \Sigma^\infty| pre(v) \subseteq K\}$;
then the {\it $\omega$-closure} of an $\omega$-language $R$ is given by
\begin{align}\label{eq:def_clo}
clo(R) := lim(pre(R)) = pre^{-1}(pre(R)) \cap \Sigma^\omega.
\end{align}
If $R = clo(R)$, we say that $R$ is {\it $\omega$-closed};
if $R = clo(R) \cap S$, we say that $R$ is {\it $\omega$-closed with respect to $S$}.
Note that $S(\bf G)$ represents a liveness assumption in the modelling of ${\bf G}$, and
in general $S({\bf G}) \subseteq lim(L({\bf G}))$; so $S(\bf G)$ itself need not be $\omega$-closed.

% the end

\subsection{Supervisory Control for DES with Infinite Behavior} \label{Subsec:SupControl}

For supervisory control, the event set $\Sigma$ is partitioned into $\Sigma_c$,
the subset of {\it controllable events} that can be disabled by an external
supervisor, and $\Sigma_{uc}$, is the subset of {\it uncontrollable events}
that cannot be prevented from occurring (i.e. $\Sigma = \Sigma_c ~\dot\cup~ \Sigma_{uc}$).
A {\it supervisory control} for $\bf G$ is any map $f: L({\bf G}) \rightarrow \Gamma$,
where $\Gamma := \{\gamma \subseteq \Sigma| \gamma \supseteq \Sigma_{u}\}$.
Then the finite and infinite closed-loop behaviors of the {\it controlled DES}
${\bf G}^f$, representing the action of the supervisor $f$ on $\bf G$,
are respectively given by
\begin{enumerate}[(a)]
\item $L({\bf G}^f)$, the $*$-language synthesized by $f$, is defined
by the following recursion:
\begin{align*}
\mbox{(i)} ~&\epsilon \in L({\bf G}^f), \\
\mbox{(ii)} ~&(\forall s \in \Sigma^*, \sigma\in \Sigma) ~s\sigma \in L({\bf G}^f)
\Leftrightarrow s \in L({\bf G}^f) ~\&~\\
            &s\sigma \in L({\bf G}) ~\&~ \sigma \in f(s);\\
\end{align*}
\item $S({\bf G}^f)$, the $w$-language synthesized by $f$, is given by
\begin{align} \label{eq:infcontrolbehaiv}
S({\bf G}^f):= lim(L({\bf G}^f)) \cap S({\bf G})
\end{align}
\end{enumerate}

The definition of $L({\bf G}^f)$ means that a string $s\sigma$ can
occur under supervision if and only if the string $s$ can occur under supervision,
and the event $\sigma$ can take place without violating either the `physical'
constraints embodied by $L({\bf G})$ or the control pattern imposed by the supervisor.
The definition of $S({\bf G}^f)$ says that an infinite string $s \in S({\bf G})$
can eventually occur if and only if it can occur in the absence of supervision
and the supervisor does not prevent the occurrence of any of its its prefixes in $pre(s)$.
Namely, $f$ exert its influence on infinite strings only through the control actions
on their finite prefixes. We say that $f: L(\bf G) \rightarrow \Gamma$ is a {\it complete}
supervisor for $\bf G$ if $L({\bf G}^f) \subseteq L({\bf G})$, and a {\it deadlock-free}
supervisor if ${\bf G}^f$ is a deadlock-free DES, i.e. $pre(S({\bf G}^f)) = L({\bf G}^f)$.

There are two classes of control requirements imposed on $\bf G$: {\it safety}
specifications describing that some conditions on $\bf G$ {\it will not} occur,
and {\it liveness} specifications requiring that some other conditions
must occur {\it eventually} \cite{Lampor77}. The safety and liveness specifications
can be specified in terms of $*$-languages and $\omega$-languages, respectively.
%Assume that the plant ${\bf G}$ as in (\ref{eq:plant}) is imposed by a
%safety specification  $E_s \subseteq \Sigma^*$, a liveness specification
%$E_l \subseteq \Sigma^\omega$;
In the following we briefly introduce the supervisory control for $\bf G$ with infinite behavior.

First, for safety specifications, consider supervisory control of
the finite behavior of $\bf G$;
it is proved \cite{GolRam87} that there exists
a complete supervisor $f^*:L({\bf G}) \rightarrow \Gamma$ that
synthesizes a $*$-language $K \subseteq L({\bf G})$ if and only if
$K$ is {\it $*$-controllable} with respect to $\bf G$
and $*$-closed with respect to $L(\bf G)$.

Formally, language $K \subseteq \Sigma^\infty$ is {\it $*$-controllable}
with respect to $\bf G$ (or $L({\bf G})$) if
\begin{align*} %\label{eq:*control}
pre(K) \Sigma_u \cap pre(L({\bf G})) \subseteq pre(K).
\end{align*}
Let $*$-language $E_s$ represent a safety specification imposed on $\bf G$,
and
\begin{align*}
\mathcal{C}^*(E_s):= \{K \subseteq L({\bf G})|~ &K \subseteq E_s ~
\text{and} \\
&\text{$K$ is $*$-controllable wrt. ${\bf G}$ and}\\
&\text{$*$-closed wrt.}~ L({\bf G})\}
\end{align*}
the set of $*$-controllable and $*$-closed sublanguages of $E_s$.
Since $*$-controllability and $*$-closure are both closed under arbitrary
set union, there exists the supremal $*$-controllable and $*$-closed sublanguage
$\sup\mathcal{C}^*(E_s)$ which may be effectively computed,
and furthermore, a complete and deadlock-free supervisor
\begin{align} \label{eq:f*}
f^*:L({\bf G}) \rightarrow \Gamma
\end{align}
synthesizing $\sup\mathcal{C}^*(E_s)$, i.e.
\begin{align*}
L({\bf G}^{f^*}) &= \sup\mathcal{C}^*(E_s )
\end{align*}
can be constructed  \cite{Wonham16a,GolRam87}. %Namely, the finite behavior of $\bf G$ under the control
%of $f^*$ is $L({\bf G}^{f^*})$ and thus the corresponding infinite behavior of $\bf G$
%is $S({\bf G}^{f^*}) = S({\bf G}) \cap lim(L({\bf G}^{f^*}))$.

Then for liveness specifications, consider supervisory control of infinite behavior
of $\bf G$; it is proved \cite[Proposition 4.5]{ThiWon94b} that there exists a complete and
deadlock-free supervisor $f^\omega:L({\bf G}) \rightarrow \Gamma$ that
synthesizes an $\omega$-language $T \subseteq S({\bf G})$ if and only if
$T$ is {\it $\omega$-controllable} with respect to $\bf G$
and $\omega$-closed with respect to $S(\bf G)$. To introduce
$\omega$-controllability, we need the concept of {\it controllability prefix}.

For an $\omega$-language $T \subseteq \Sigma^\omega$, its {\it controllability prefix} is given by
\begin{align*}
pre_{\bf G}(T) := \{t \in pre(T)|(\exists T' \subseteq T/t) ~[T' \neq\emptyset
~\text{is}& \\
*\text{-controllable wrt.}~ &L({\bf G})/t~\notag\\
 \text{and} ~\omega\text{-closed wrt.} ~ &S({\bf G})/t~]\}
\end{align*}
where $T/t := \{s \in \Sigma^\infty| ts \in T\}$, and $L({\bf G})/t$
and $S({\bf G})/t$ are defined similarly.

Now, we define that $T$ is {\it $\omega$-controllable} with respect to $\bf G$ if
\begin{align*}
(\textnormal{i}) ~&\text{$T$ is $*$-controllable with respect to $\bf G$;}\\
(\textnormal{ii}) ~&\text{$pre(T) = pre_{\bf G}(T)$.}
\end{align*}
Note that $\omega$-controllable and $\omega$-closed languages have different closure properties under union and
intersection. Specifically, $\omega$-controllability is preserved under
arbitrary unions but not intersections, while $\omega$-closure is preserved
under arbitrary intersections but not unions. It is therefore convenient
to define, below, the separate language classes:
\begin{align*}
\mathcal{C}^\omega(E_l):= \{T \subseteq S({\bf G})|~ &T \subseteq E_l ~
\text{and} \\
&\text{$T$ is $\omega$-controllable wrt.}~ {\bf G}\} \\
\mathcal{F}^\omega(A):= \{T \subseteq S({\bf G})|~ &A \subseteq T ~
\text{and} \\
&\text{$T$ is $\omega$-closed wrt.}~ S({\bf G})\}
\end{align*}
where $E_l$ is an $\omega$-language representing the {\it maximal legal} specification and $A$ is
also an $\omega$-language but representing the {\it minimal acceptable} specification.
Due to the closure property of
$\omega$-controllability and $\omega$-closure described above, there exists
\cite[Proposition 5.2]{ThiWon94b} the unique supremal $\omega$-controllable sublanguage
$\sup\mathcal{C}^\omega(E_l)$, given by
\begin{align*}
\sup\mathcal{C}^\omega(E_l) := lim(pre_{\bf G}(E_l)) \cap E_l
\end{align*}
and the unique infimal $\omega$-closed superlanguage $\inf\mathcal{F}^\omega(A)$, given by
\begin{align*}
\inf\mathcal{F}^\omega(A) := clo(A) \cap S({\bf G}).
\end{align*}
Furthermore, it is proved \cite[Theorem 5.3]{ThiWon94b} that there exists a
$\omega$-controllable and $\omega$-closed language $T$ such that
%\begin{align} \label{eq:scpw}
$A \subset T \subseteq E_l$
%\end{align}
if and only if
\begin{align} \label{eq:control_exist}
\inf\mathcal{F}^\omega(A) \subseteq \sup\mathcal{C}^\omega(E_l)
\end{align}
and if exists, a complete and deadlock-free supervisor
\begin{align} \label{eq:f_w}
f^\omega : L({\bf G}) \rightarrow \Gamma
\end{align}
synthesizing such $T$, i.e.
\begin{align}
A \subset T = S({\bf G}^{f^\omega}) \subseteq E_l \label{eq:sub1:scpw}\\
pre(S({\bf G}^{f^\omega})) = L({\bf G}^{f^\omega}) \label{eq:sub2:scpw}
\end{align}
can be constructed according to the procedure described in Appendix~\ref{Append:SupSynth}.

% the end

\section{Problem Formulation} \label{Sec:ProbForm}

%\input{Subsec_SupSynth.tex}

%\subsection{Problem Formulation} \label{Subsec:ProbForm}

Let $\bf G$ as in (\ref{eq:plant}) be the plant to be controlled,
$E_s$ the safety specification, $E_l$ the maximal legal liveness specification,
and $A$ the minimal acceptable liveness specification. To synthesize supervisors
for these specifications, our approach is in a simple but natural way: first synthesize
a supervisor $f^*$ for the safety specification; then treat the closed-loop
behavior of $\bf G$ controlled by $f^*$ as the new plant to be controlled, and synthesize
another supervisor $f^\omega$ for the liveness specifications.
By this approach, the supervisors $f^*$ and $f^\omega$ work conjunctively,
without conflicts, because the controlled behavior of $f^*$ is the plant behavior of $f^\omega$ and thus
a controllable event that has been disabled by $f^*$ need
not be disabled by $f^\omega$ again.

%The behavior of $\bf G$ under the control of $f^*$ and $f^\omega$ will
%be verified to satisfy both the safety specification $E_s$ and the maximal legal liveness specification
%$E_l$, and contain the minimally

First, for the safety specification $E_s$, we synthesize as in (\ref{eq:f*})
a complete supervisor $f^*:L({\bf G})\rightarrow \Gamma$
such that the finite behavior of $\bf G$ under the control of $f^*$, denoted by ${\bf G}^{f^*}$,
satisfies
\begin{align*}
L({\bf G}^{f^*}) &= \sup\mathcal{C}^*(E_s ).
%S({\bf G}^{f^*}) &= S({\bf G}) \cap lim(L({\bf G}^{f^*}))
\end{align*}
According to (\ref{eq:infcontrolbehaiv}), the infinite controlled behavior of ${\bf G}$
is $S({\bf G}^{f^*}) = S({\bf G}) \cap lim(L({\bf G}^{f^*}))$;
${\bf G}^{f^*}$ can be represented by a deterministic B\"uchi automaton
$(X^*, \Sigma, \xi^*, x_0^*, \mathcal{B}_{X^*})$ constructed according to
\begin{enumerate}[(i)]

\item Select $x_0^* \in X^*$ corresponding to $q_0 \in Q$.

\item $\xi^* : X^*\times \Sigma \rightarrow X^*$ with $\xi^*(x^*,\sigma) = x'^{*}$
if there exists $s \in \Sigma^*$ such that $\xi^*(x_0^*, s) = x^*$, $\delta(q_0,s\sigma)!$, and $\sigma \in f^*(s)$.

\item $\mathcal{B}_{X^*} := \{x^* \in X^*| (\exists s \in \Sigma^*)~ \xi^*(x_0^*, s) = x^*, \delta(q_0, s) \in \mathcal{B}_Q\}$.
\end{enumerate}

Let
\begin{align}\label{eq:monosup*}
{\bf SUP}^* := (X^*, \Sigma, \xi^*, x_0^*).
\end{align}
Namely ${\bf SUP}^*$ has the same transition structure and thus same finite behavior as ${\bf G}^{f^*}$,
i.e.
%\begin{align}
$L({\bf SUP}^*) = L({\bf G}^{f^*})$.
%\end{align}
Then ${\bf SUP}^*$ is an {\it implementation} \cite{Wonham16a}
of the supervisor $f^*$, i.e.
\begin{align*}
L({\bf G}) \cap L({\bf SUP}^*) &= L({\bf G}^{f^*}) %\label{eq:sub1:sup*}\\
%S({\bf G}) \cap lim(L({\bf SUP}^*)) &= S({\bf G}^{f^*}) %\label{eq:sub2:sup*}
\end{align*}
Since $L({\bf SUP}^*) = L({\bf G}^{f^*})$, it also infers that
%\begin{align}
$S({\bf G}^{f^*}) = S({\bf G}) \cap lim(L({\bf SUP}^*))$. %\label{eq:def_SUP_*}
%\end{align}

%${\bf SUP}^*$ is referred as the (automaton-based) monolithic supervisor for $\bf G$.

Second, we consider the supervisor synthesis for the liveness specifications $E_l$ and $A$.
At this step, we treat ${\bf G}^{f^*}$ as the new plant to be controlled, and synthesize
as in (\ref{eq:f_w}) a complete and deadlock-free supervisor $f^\omega:L({\bf G}^{f^*})\rightarrow \Gamma$
given by
\begin{equation} \label{eq:def_fw}
%\begin{split}
f^\omega(l) :=
\left\{
   \begin{array}{ll}
      f_0^\omega(l) & \text{if}~ l \in pre(A), \\
      f_k^\omega(l/k) & \text{if}~ l \in k ~pre(E_k'), k\in M \\
      \text{undefined} & \text{otherwise}
   \end{array}
\right.
%\end{split}
\end{equation}
where $M$ is the set of all elements of $pre(\sup\mathcal{C}^\omega(E_l))\setminus pre(\inf\mathcal{F}^\omega(A))$ of minimal length,
and $E_k'$ is the sublanguage of $\sup\mathcal{C}^\omega(E_l)/k$ synthesized by $f_k^\omega$.
%The supervisor $f^\omega$ is comprised of a complete and deadlock-free supervisor
%$f_0^\omega: \Sigma^* \rightarrow \Gamma$, which synthesizes the $\omega$-closed language $\inf\mathcal{F}^\omega(A)$, and
%a set of complete and deadlock-free supervisors $f_k^\omega: \Sigma^* \rightarrow \Gamma$ one for
%each element $k \in M$, which synthesizes some $\omega$-sublanguage $E_k'$.
Under the supervision of $f^\omega$, the infinite controlled behavior of ${\bf G}^{f^*}$,
denoted by ${\bf G}^{f^*\wedge f^\omega}$ ($f^*$ and $f^\omega$ work conjunctively, i.e.
a controllable event will be disabled if it is disabled by
any one of $f^*$ and $f^\omega$), satisfies:
\begin{align*}
&A \subseteq S({\bf G}^{f^*\wedge f^\omega}) \subseteq E_l \\ %\label{eq:sub1:scpw}\\
&pre(S({\bf G}^{f^*\wedge f^\omega})) = L({\bf G}^{f^*\wedge f^\omega}). %\label{eq:sub2:scpw}
\end{align*}
${\bf G}^{f^*\wedge f^\omega}$ can be represented by a deterministic B\"uchi automaton
$(X^\omega, \Sigma, \xi^\omega, x_0^\omega, \mathcal{B}_{X^\omega})$ constructed by:
\begin{enumerate}[(i)]

\item Select $x_0^\omega \in X^\omega$ corresponds to $x_0^* \in X^*$.

\item $\xi^\omega : X^\omega\times \Sigma \rightarrow X^\omega$ with $\xi^\omega(x^\omega,\sigma) = {x'}^\omega$
if there exists $s \in \Sigma^*$ such that $\xi^\omega(x_0^\omega, s) = x^\omega$, $\xi^*(x_0^*,s\sigma)!$, and $\sigma \in f^\omega(s)$.

\item $\mathcal{B}_{X^\omega} := \{x^\omega \in X^\omega| (\exists s \in \Sigma^*)~ \xi^\omega(x_0^\omega, s) = x^\omega, \xi(x_0^*, s) \in \mathcal{B}_{X^*}\}$.
\end{enumerate}
%The finite and infinite behaviors of ${\bf G}^{f^\omega}$ are:
%\begin{align}
%L({\bf G}^{f^\omega}) &:= pre(\inf\mathcal{F}^\omega(A)) \cup \bigcup_{k\in M}k~ pre(E_k') \label{eq:sub1:implement}\\
%S({\bf G}^{f^\omega}) &:= \inf\mathcal{F}^\omega(A) \cup \bigcup_{k\in M}k~ E_k' \label{eq:sub2:implement}
%\end{align}

The supervisor $f^\omega:\Sigma^*\rightarrow \Gamma$ exercises its control action
depending on its observation on finite strings in $\Sigma^*$, and thus $f^\omega$ also
can be implemented by a *-automaton. Let
\begin{align}\label{eq:monosup_w}
{\bf SUP}^\omega := (X^\omega, \Sigma, \xi^\omega, x_0^\omega).
\end{align}
Namely ${\bf SUP}^\omega$ has the same transition structure and
thus same finite behavior as ${\bf G}^{f^*\wedge f^\omega}$,
i.e.
%\begin{align}
$L({\bf SUP}^\omega) = L({\bf G}^{f^*\wedge f^\omega})$.
%\end{align}
Then ${\bf SUP}^\omega$ is an {\it implementation} of the supervisor $f^\omega$,
i.e.
\begin{align}
%L({\bf G}^{f^*}) \cap L({\bf SUP}^\omega) &= L({\bf G}^{f^*\wedge f^\omega}) \\%\label{eq:sub1:implement}\\
S({\bf G}^{f^*}) \cap lim(L({\bf SUP}^\omega)) &= S({\bf G}^{f^*\wedge f^\omega}).\label{eq:def_SUP_w}
\end{align}

Note that ${\bf SUP}^\omega$ also influences the finite controlled behavior
of $\bf G$, thus $L({\bf G}^{f^*\wedge f^\omega})$ and $S({\bf G}^{f^*\wedge f^\omega})$ represent
respectively the finite and infinite controlled behavior of $\bf G$
under the control of ${\bf SUP}^*$ and ${\bf SUP}^\omega$, i.e.
\begin{align}
&L({\bf G}^{f^*\wedge f^\omega}) = L({\bf G}) \cap L({\bf SUP}^*) \cap L({\bf SUP}^\omega)  \label{eq:sub1:implement}\\
&S({\bf G}^{f^*\wedge f^\omega}) = S({\bf G}) \cap lim(L({\bf SUP}^*)) \cap lim(L({\bf SUP}^\omega)). \label{eq:sub2:implement}
\end{align}

It is easily verified that the finite controlled behavior of ${\bf G}$ satisfies the safety specification,
i.e. \[L({\bf G}^{f^*\wedge f^\omega}) \subseteq E_s,\] and the infinite controlled behavior
fits into the range of liveness specifications $E_l$ and $A$, i.e. \[A \subseteq S({\bf G}^{f^*\wedge f^\omega}) \subseteq E_l.\]
The supervisor ${\bf SUP}^*$ is constructed for satisfying the safety specification and thus we
refer it as the {\it safety supervisor} for $\bf G$; while ${\bf SUP}^\omega$ is constructed for
the liveness specifications and thus we refer it as the {\it liveness supervisor} for $\bf G$.
Throughout this paper, we assume that $S({\bf G}^{f^*\wedge f^\omega})\neq \emptyset$ and
thus $L({\bf G}^{f^*\wedge f^\omega})\neq \emptyset$.

\begin{figure}[!t]
\centering
    \includegraphics[scale = 0.8]{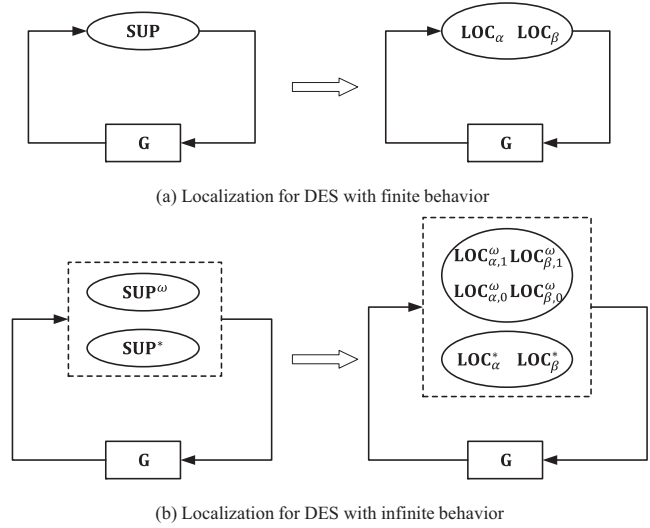}
\caption{Supervisor localization example for illustration: let $\Sigma_c = \{\alpha,\beta\}$.
For DES $\bf G$ with finite behavior as in (a), the monolithic supervisor $\bf SUP$ is decomposed into
two local controllers ${\bf LOC}_\alpha$ and ${\bf LOC}_\beta$ for controllable events $\alpha$ and
$\beta$ respectively. For DES $\bf G$ with infinite behavior as in (b), there are two supervisors ${\bf SUP}^*$
and ${\bf SUP}^\omega$ constructed for satisfying safety specification and liveness specifications respectively.
The localization procedure decomposes ${\bf SUP}^*$ into two local controllers ${\bf LOC}_\alpha^*$ and ${\bf LOC}_\beta^*$,
and decomposes ${\bf SUP}^\omega$ into local controllers ${\bf LOC}_{\alpha,0}^*$ and ${\bf LOC}_{\alpha,1}^*$
 for $\alpha$, and ${\bf LOC}_{\beta,0}^\omega$ and ${\bf LOC}_{\beta,1}^\omega$ for $\beta$.} \label{fig:localization}
\end{figure}

The control action of ${\bf SUP}^*$ and ${\bf SUP}^\omega$ are both to enable/disable controllable
events; thus the localizations of ${\bf SUP}^*$ and ${\bf SUP}^\omega$ are similar to
that of the monolithic supervisor $\bf SUP$ in \cite{CaiWon10a}.
The differences are illustrated in Fig.~\ref{fig:localization}. First, the localization
of $\bf SUP$ generate one local controller for each controllable event. However, the present localization
procedure may generate multiple local controllers for one controllable event, because
an event may be disabled/enabled by both ${\bf SUP}^*$ and ${\bf SUP}^\omega$.
Second, the localization of ${\bf SUP}^*$ is similar to that of $\bf SUP$
in \cite{CaiWon10a}, however, the localization of ${\bf SUP}^\omega$
is particular: according to whether or not $s \in pre(A)$ (see (\ref{eq:def_fw}) for the definition
of $f^\omega$), there are two types of supervisors
included in $f^\omega$: $f_0^{\omega}$ defined on the strings $s \in pre(A)$
and $f_k^\omega$ defined on the rest of the strings in $L({\bf G})$, thus the localization of ${\bf SUP}^\omega$
can be divided into two parts and consequently, we will get two local controllers
for each controllable event.

\begin{remark}
We remark here that the localization of the control actions after string $s \notin pre(A)$
is treated as a whole, but not divided corresponding to each $f_k^\omega$ ($k \in M$).
The reason is as follows. First, to localize the control actions after each string $k$,
we need to find in $L({\bf SUP}^\omega)$ the language $E_k'$ synthesized by $f_k^\omega$,
which will increase the time complexity of the overall algorithm. Second, the number
of local controllers will increase with the states number of ${\bf SUP}^\omega$.
In our current setting, all the controlled behavior synthesized by $f_k^\omega$ are
contained in $L({\bf SUP}^\omega)$, thus we don't have to find each $E_k'$;
consequently for each controllable event, ${\bf SUP}^\omega$ will be
constantly decomposed into two local controllers: one corresponding to $f_0^\omega$ and
the other to all $f_k^\omega$.
%Third, according to whether or not $s \in pre(A)$, we could convert ${\bf SUP}^\omega$
%into an equivalent supervisor ${\bf ESUP}^\omega$ (i.e. $L({\bf ESUP}^\omega) = L({\bf SUP}^\omega)$)
%of which the state set can be partitioned into two sets: string $s \in pre(A)$ can only visit
%the states in one of the sets. With this partition, the localization presented below
\end{remark}

\begin{remark}
Note that it is also possible to construct a monolithic supervisor $\bf SUP$ that synthesizes
the controlled behavior $L({\bf G}^{f^*\wedge f^\omega})$, i.e. $\bf SUP$ is control equivalent
to ${\bf SUP}^*$ and ${\bf SUP}^\omega$. In that case, by applying the localization procedure
in \cite{CaiWon10a}, we may get for each controllable event a local controller.
In general, this local controller will have more states than the local controllers constructed by
our new localization procedures,
as will be demonstrated in the example of Small Factory in Section~\ref{Sec:CaseStudy}. The reason
is that either ${\bf SUP}^*$, or ${\bf SUP}^\omega$, disables controllable events on part of
the strings in $L({\bf G})$: the plant of ${\bf SUP}^*$ is $\bf G$ and the plant of ${\bf SUP}^\omega$ is ${\bf G}^{f^*}$.
\end{remark}

Due to the above features specific to ${\bf SUP}^*$ and ${\bf SUP}^\omega$, we have
different types of local controllers for each controllable event $\alpha \in \Sigma_c$.
First, we say that a *-automaton
\begin{equation*}
{\bf LOC}_\alpha^* =
(Y_\alpha^*,\Sigma,\eta_\alpha^*,y_{0,\alpha}^*)
\end{equation*}
is a {\it safety local controller} for $\alpha$ if
${\bf LOC}_\alpha^*$ enables/disables event $\alpha$ (and only
$\alpha$) consistently with ${\bf SUP}^*$, which means that for all $s \in \Sigma^*$ there holds
\begin{align} \label{eq:loc_*}
s\alpha \in L({{\bf LOC}_\alpha^*}),\ s\alpha \in L({\bf G}),~
s \in L({\bf SUP}^*) \notag\\
\Leftrightarrow ~s\alpha \in L({\bf SUP}^*)
\end{align}

Second, for all the strings $s \in L({\bf G}^{f^*})$, we divide them into two parts:
$C_1 = pre(A)$ and $C_2 = L({\bf G}^{f^*})\setminus pre(A)$.
For each part $C_n$ ($n = 1,2$), we say that a *-automaton
\begin{equation*}
{\bf LOC}_{\alpha,n}^\omega =
(Y_{\alpha,n}^\omega,\Sigma,\eta_{\alpha,n}^\omega,y_{0,\alpha,n}^\omega),
\end{equation*}
is a {\it liveness local controller} for $\alpha$ if
${\bf LOC}_{\alpha,n}^\omega$ enables/disables event $\alpha$ (and only
$\alpha$) occurred at string $s \in C_n$ consistently with ${\bf SUP}^\omega$,
which means that for all $s \in C_n$ there holds
\begin{align} \label{eq:loc_w}
s\alpha \in L({{\bf LOC}_{\alpha,n}^*}),\ s\alpha \in L({\bf G}^{f^*}),~
s \in L({\bf SUP}^\omega) \notag\\
\Leftrightarrow ~s\alpha \in L({\bf SUP}^\omega)
\end{align}

We now formulate the {\it Supervisor Localization Problem} for DES with infinite behavior:

%\begin{enumerate}[(i)]
%\item
Construct a set of safety local controllers $\{{\bf LOC}_\alpha^*\ |\ \alpha \in \Sigma_c\}$,
a set of liveness local controllers $\{{\bf LOC}_{\alpha,n}^\omega\ |\ \alpha \in \Sigma_c, n = 1, 2\}$
such that their collective controlled behaviors are equivalent to those of
supervisors ${\bf SUP}^*$ and ${\bf SUP}^\omega$ with respect to ${\bf G}$, i.e.
\begin{align*}{\scriptsize}
   L({\bf G}) &\cap\Big(\mathop \bigcap\limits_{\alpha \in \Sigma_{c}} L({\bf LOC}_{\alpha}^*) \Big) \notag \\
              &\cap \Big(\mathop \bigcap\limits_{\alpha \in \Sigma_{c}, n = 1, 2} L({\bf LOC}_{\alpha,n}^\omega) \Big)
               =~L({\bf G}^{f^* \wedge f^\omega}) \\ %\label{eq:sub1_equiv}
   S({\bf G}) &\cap \Big(\mathop \bigcap\limits_{\alpha \in \Sigma_{c}} lim(L({\bf LOC}_{\alpha}^*)) \Big) \notag \\
              &\cap \Big(\mathop \bigcap\limits_{\alpha \in \Sigma_{c}, n = 1,2} lim(L({\bf LOC}_{\alpha,n}^\omega)) \Big)
               = ~S({\bf G}^{f^* \wedge f^\omega})   %\label{eq:sub2_equiv}
\end{align*}
where $L({\bf G}^{f^*\wedge f^\omega})$ and $S({\bf G}^{f^*\wedge f^\omega})$
%\begin{align}
%&L({\bf G}^{f^*\wedge f^\omega}) = L({\bf G}) \cap L({\bf SUP}^*) \cap L({\bf SUP}^\omega)  \label{eq:sub1:implement}\\
%&S({\bf G}^{f^*\wedge f^\omega}) = S({\bf G}) \cap lim(L({\bf SUP}^*)) \cap lim(L({\bf SUP}^\omega)) \label{eq:sub2:implement}
%\end{align}
respectively represent the finite and infinite controlled behaviors of $\bf G$
under the control of ${\bf SUP}^*$ and ${\bf SUP}^\omega$ (as in (\ref{eq:sub1:implement}) and (\ref{eq:sub2:implement})).

Having obtained these local controllers for individual controllable event, for the plant consisting of multiple components,
we can allocate each controller to the agent(s) owning the corresponding controllable event.
Thereby we build for a multi-agent DES with infinite behavior a nonblocking distributed control
architecture.

%% the end

%\input{Subsec_ProductSys.tex}

\section{Supervisor Localization Procedure} \label{Sec:suploc}

We solve the Supervisor Localization Problem for DES with infinite behavior
by extending the localization procedure proposed in \cite{CaiWon10a}.
In particular, localization of ${\bf SUP}^\omega$ will be divided into two cases
by considering the control action of $f_0^\omega$ and those of $f_k^\omega$ separately,
for which we introduce new definition of control consistency relation.

Given a DES plant ${\bf G} = (Q,\Sigma,\delta, q_0, \mathcal{B}_Q)$ (as in (\ref{eq:plant}))
with a safety supervisor ${\bf SUP}^* = (X^*,\Sigma,\xi^*,x_0^*)$ and a
liveness supervisor ${\bf SUP}^\omega$, we present the localization of ${\bf SUP}^\omega$
(with new control consistency concept) and that of ${\bf SUP}^*$ in the sequel.

% the end

\subsection{Localization of ${\bf SUP}^\omega$} \label{Subsec:SupLocLive}

As mentioned in Section~\ref{Sec:prelin}, an infinite string $s$
can eventually occur if and only if it can occur in the absence of supervision
and the supervisor does not prevent the occurrence of any of its its prefixes in $pre(s)$.
In other words, the supervisor ${\bf SUP}^\omega$ (implementation of $f^\omega$)
exerts its influence on infinite strings only through the control actions
on their finite prefixes. So, the localization procedure for ${\bf SUP}^\omega$
is to decompose the control actions on the finite strings $s \in L({\bf G}^{f^*})$
(the plant of ${\bf SUP}^\omega$), and as in \cite{CaiWon10a}, the control equivalence of
finite behaviors will be guaranteed by the localization procedure.
The control equivalence of infinite behaviors, however, will be derived by the following Lemma
once the equivalence of finite behaviors were confirmed.

\begin{lem} \label{lem:equ_lim}
Let $A, B, C \subseteq \Sigma^*$ be arbitrary $*$-languages, then we have
\[A \cap B = C \Rightarrow lim(A) \cap lim (B) = lim(C)\]
where the operator $lim$ is defined in (\ref{eq:def_lim}).
\end{lem}

\noindent {\it Proof}: Recall that (see (\ref{eq:def_lim}))
$lim(A) = pre^{-1}(A) \cap \Sigma^\omega := \{t \in \Sigma^\omega | pre(t) \subseteq A\}$.

($\supseteq$) By the above definition and $C\subseteq A \cap B$, we have $lim(C) \subseteq lim(A)$ and
$lim(C) \subseteq lim(B)$. So $lim(C) \subseteq lim(A) \cap lim(B)$.

($\subseteq$) Let $s \in lim(A) \cap lim (B)$. Then $s \in lim(A)$, and thus
$pre(s) \subseteq A$; by the same reason, $pre(s) \subseteq B$. Hence
$pre(s) \subseteq A \cap B = C$, and thus $s \in lim(C)$, which completes the proof. \hfill $\square$

\vspace{1em}

The control action of ${\bf SUP}^\omega$ is to enable or disable controllable events
in $\Sigma_c$ at strings $s \in L({\bf G}^{f^*})$. As in (\ref{eq:def_fw}), the control action
after a string $s$ is divided into two cases: according to the strings $s \in C_1$ or $C_2$. Thus,
for each controllable event $\alpha$, we propose to decompose ${\bf SUP}^\omega$ into two
local controllers, one responsible for disabling $\alpha$ at strings $s \in C_n$, $n = 1$ or $2$;
in other words, the local controller corresponding to $C_n$ will not disable $\alpha$
at the string $t \in C_m (m =1, or ~2, m\neq n)$, even $\alpha$ is disabled
by ${\bf SUP}^\omega$ (although it will be disabled by the local controller corresponding to $C_m$).
Consequently, the two local controllers generally have states number no more than that obtained by
considering the disablement after all the strings in $L({\bf G}^{f^*})$.

Fix an arbitrary controllable event $\alpha \in \Sigma_{c}$ and one part of the language $C_n$, $n = 1,2$
(recall that $C_1 = pre(A)$ and $C_2 = L({\bf G}^{f^*})\setminus pre(A)$).
The control action of ${\bf SUP}^\omega$ is captured by the following two functions.
First define $E_\alpha^\omega: X^\omega \rightarrow \{1,0\}$ according to
\begin{equation}\label{eq:E_w}
E_\alpha^\omega(x^\omega) = 1~\text{iff}~\xi^\omega(x^\omega,\alpha)!
\end{equation}
So $E_\alpha^\omega(x^\omega) = 1$ means that $\alpha$ is defined at state $x^\omega$ in
${\bf SUP}^\omega$. Next define $D_{\alpha,n}^\omega: X^\omega\rightarrow\{1,0\}$ according to
$D_{\alpha,n}^\omega(x^\omega) = 1$ iff
\begin{align}\label{eq:D_w0}
\neg\xi^\omega(x^\omega,\alpha)! \ ~\&~\ &(\exists s\in C_n) \notag\\
&\big(\xi^\omega(x_0^\omega,s)=x^\omega ~\&~\xi^*(x_0^*, s\alpha)!  \big)
\end{align}
Thus $D_{\alpha,n}^\omega(x) = 1$ means that $\alpha$ must be disabled at $x$ arrived by strings
$s \in C_n$ consistently with the supervisor ${\bf SUP}^\omega$ (i.e. $\alpha$ is disabled at $x$ in ${\bf SUP}^\omega$
but is defined at some state in the plant ${\bf G}^{f^*}$ corresponding to $x$ via string
$s \in C_n$). Note that here the plant is ${\bf G}^{f^*}$, not ${\bf G}$,
because as in Section~\ref{Sec:ProbForm} when synthesizing the supervisor ${\bf SUP}^\omega$, ${\bf G}^{f^*}$ is considered
as the plant to be controlled.

\begin{figure}[!t]
\centering
    \includegraphics[scale = 0.8]{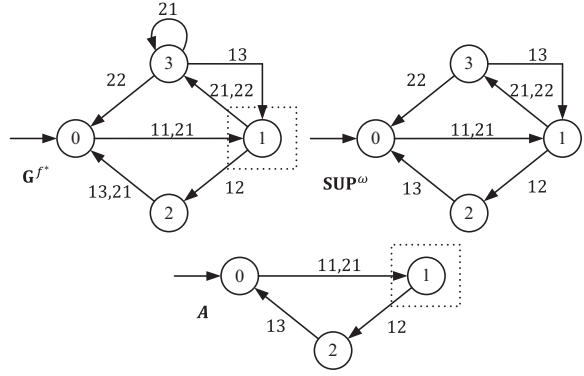}
\caption{Example: Plant ${\bf G}^{f^*}$ (B\"uchi automaton), supervisor ${\bf SUP}^\omega$ ($*$-automaton)
and B\"uchi automaton $\bf A$ representing the minimal acceptable liveness specification $A$.
Notations: a circle with right input arrow $\rightarrow$ denotes the initial state, and a circle in dotted box
denotes that this state is an element of the B\"uchi acceptance criterion; we shall use these notations throughout
this report.} \label{fig:Examp_GSUP}
\end{figure}

The function $D_{\alpha,n}^\omega$ differs from that in \cite{CaiWon10a} in the range of
strings $s$: here $D_{\alpha,n}^\omega(x^\omega) = 1$ only when $x^\omega$ can be arrived by a string $s \in C_n$.
For illustration, consider the example in Fig.~\ref{fig:Examp_GSUP}:  $D_{21,1}^{\omega}(2) = 1$ because state
$2$ can be reached by string $s = 11.12 \in C_1 = pre(A)$; however, $D_{21,1}^{\omega}(3) = 0$, by the
reason that none of the strings in $C_1$ can reach state $3$.

Based on (\ref{eq:E_w}) and (\ref{eq:D_w0}), we define the following binary
relation $\mathcal{R}_{\alpha,n}^\omega \subseteq X^\omega\times X^\omega$, called {\it
control consistency} with respect to controllable event $\alpha$ (cf.
\cite{CaiWon10a}), according to $(x^\omega,x'^\omega)\in
\mathcal{R}_{\alpha,n}$ iff
\begin{align} \label{eq:consist_w}
E_\alpha(x^\omega)\cdot D_{\alpha,n}^\omega(x'^\omega) = 0 = E_\alpha(x'^\omega)\cdot D_{\alpha,n}(x^\omega)
\end{align}

Thus a pair of states $(x^\omega,x'^\omega)$ in ${\bf SUP}^\omega$ satisfies $(x^\omega,x'^\omega) \in
\mathcal{R}_{\alpha,n}^\omega$ if  event $\alpha$ is
defined at one state, but not disabled at the other.
It is easily verified as in \cite{CaiWon10a} that $\mathcal{R}_{\alpha,n}^\omega$ is generally not transitive,
thus not an equivalence relation. Now let
$I^\omega$ be some index set, and $\mathcal{C}_{\alpha,n}^\omega = \{X_i^\omega\subseteq
X^\omega|i \in I^\omega\}$ a cover on $X^\omega$. $\mathcal{C}_{\alpha,n}^\omega$ is a {\it control cover} with
respect to $\alpha$ if
\begin{align*}
\mbox{(\rmnum{1})}~&(\forall i\in I^\omega, \forall x^\omega, x'^\omega \in X_i^\omega) (x^\omega,x'^\omega) \in \mathcal{R}_{\alpha,n}^\omega, \notag\\
\mbox{(\rmnum{2})}~&(\forall i\in I^\omega, \forall \sigma \in \Sigma)\Big[(\exists x^\omega\in X_i^\omega)\xi^\omega(x^\omega,\sigma)! \Rightarrow \\
&\big((\exists j\in I^\omega)(\forall x'^\omega\in X_i^\omega)\xi^\omega(x'^\omega,\sigma)! \Rightarrow
\xi^\omega(x'^\omega,\sigma)\in X_j^\omega\big)\Big]. \notag
\end{align*}
We call $\mathcal{C}_{\alpha,n}^\omega$
a {\it control congruence} if it happens to
be a partition on $X^\omega$, namely its cells
are pairwise disjoint.

Having defined a preemption cover $\mathcal{C}_{\alpha,n}^\omega$ on $X^\omega$,
we construct a local controller ${\bf LOC}_{\alpha,n}^\omega =
(Y_{\alpha,n}^\omega, \Sigma, \zeta_{\alpha,n}^\omega,\\ y_{0,\alpha,n}^\omega)$
for the controllable event $\alpha$ as follows.

\begin{enumerate}[(i)]

\item The state set is $Y_{\alpha,n}^\omega := I^\omega$, with each state $y^\omega
\in Y_{\alpha,n}^\omega$ being a cell $X_i^\omega$ of the cover
$\mathcal{C}_{\alpha,n}^\omega$. In particular, the initial state
$y_{0,\alpha,n}^\omega$ is a cell $X_{i,0}^\omega$ where $x_0^\omega$ belongs, i.e. $x_0^\omega \in
X_{i,0}^\omega$.

\item Define the transition function $\zeta_{\alpha,n}^\omega:I^\omega \times\Sigma \rightarrow I^\omega$ over
the entire event set $\Sigma$ by $\zeta_{\alpha,n}^\omega(i,\sigma)=j$ if
\begin{align*} %\label{e277}
&(\exists x^\omega\in X_i^\omega)~\xi^\omega(x^\omega,\sigma)\in X_j^\omega ~\mbox{and}~~\notag\\
&~~~~~~(\forall x'^\omega\in X_i^\omega)\big[\xi^\omega(x'^\omega,\sigma)! \Rightarrow
\xi^\omega(x'^\omega,\sigma)\in X_j^\omega\big].
\end{align*}

\end{enumerate}

Similar to Lemma 2 in \cite{ZhangCW17}, it is easily verified that
${\bf LOC}_{\alpha,n}^\omega$ constructed above is a liveness local controller for $\alpha$,
i.e. condition (\ref{eq:loc_w}) holds for all $s \in C_n$.
By the above two procedures, for one controllable event $\alpha$, we get two liveness local
controllers: ${\bf LOC}_{\alpha,1}^\omega$ responsible for the disablement at strings $s\in C_1 = pre(A)$ and
${\bf LOC}_{\alpha,2}^\omega$ for the disablement at strings $s\in C_2 = L({\bf G}^{f^*})\setminus pre(A)$.

\begin{figure}[!t]
\centering
    \includegraphics[scale = 0.8]{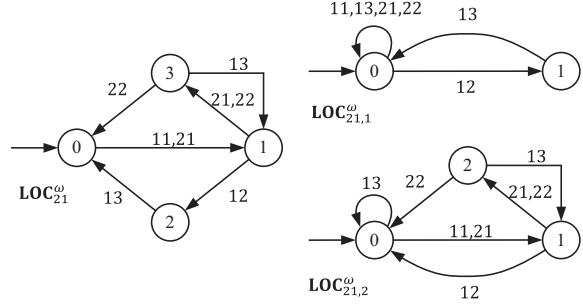}
\caption{Example: Local controllers for event $21$ of DES example in Fig.~\ref{fig:Examp_GSUP}} \label{fig:Examp_LOC}
\end{figure}

For the example in Fig.~\ref{fig:Examp_GSUP}, we get two liveness local controllers
${\bf LOC}_{21,1}^\omega$ and ${\bf LOC}_{21,2}^\omega$ for event $21$, as displayed in Fig.~\ref{fig:Examp_LOC}.
In the transition diagram of ${\bf LOC}_{21,1}^\omega$, state 0 corresponds to cell $\{0,1,3\}$
of the control cover $\mathcal{C}_{21,1}^\omega = \{\{0,1,3\},\{2\}\}$ and state 1 corresponds to
cell $\{2\}$; in ${\bf LOC}_{21,2}^\omega$, state 0 corresponds to cell $\{0,2\}$
of the control cover $\mathcal{C}_{21,2}^\omega = \{\{0,2\},\{1\},\{3\}\}$, state 1 corresponds to
cell $\{1\}$, and state 2 corresponds to cell $\{3\}$.
However, if consider the disablement at all the strings in $L({\bf G}^{f^*})$ together,
the supervisor ${\bf SUP}^\omega$ is not localizable and thus
we get a $4$-states local controller ${\bf LOC}_{21}^\omega$, which has more states
than any of ${\bf LOC}_{21,1}^\omega$ and ${\bf LOC}_{21,2}^\omega$.

%% the end

\subsection{Localization of ${\bf SUP}^*$} \label{Subsec:SupLocSafe}

The localization of ${\bf SUP}^*$ is similar to that of $\bf SUP$
in \cite{CaiWon10a}, namely, the disablement at all strings in
$L({\bf G})$ are considered. The control action of ${\bf SUP}^*$
is captured by the following two functions.

Fix an arbitrary controllable event $\alpha \in \Sigma_{c}$. First
define $E_\alpha^*: X^* \rightarrow \{1,0\}$ according to
\begin{equation}\label{eq:E_*}
E_\alpha^*(x^*) = 1~\text{iff}~\xi^*(x^*,\alpha)!
\end{equation}
So $E_\alpha^*(x^*) = 1$ means that $\alpha$ is defined at state $x^*$ in
${\bf SUP}^*$. Next define $D_\alpha^*: X^* \rightarrow\{1,0\}$ according to
$D_\alpha^*(x^*) = 1$ iff
\begin{eqnarray}\label{eq:D_*}
\neg\xi^*(x^*,\alpha)! \ ~\&~\ (\exists s\in \Sigma^*)\left(\xi^*(x_0^*,s)=x^*
~\&~\delta(q_0, s\alpha)! \right)
\end{eqnarray}
Thus $D_\alpha^*(x^*) = 1$ means that $\alpha$ must be disabled at $x^*$
(i.e. $\alpha$ is disabled at $x^*$ in ${\bf SUP}^*$ but is defined at some
state in the plant $\textbf{G}$ corresponding to $x^*$ via string
$s$).

With new definition of $D_{\alpha}^*$, we get new definitions of
control consistency relation $\mathcal{R}_{\alpha}^*$ and control cover $\mathcal{C}_{\alpha}^*$,
and then by the rules (i)-(ii) for constructing liveness local controller replaced with the new definitions,
we construct a new local controller
${\bf LOC}_{\alpha}^* = (Y_{\alpha}^*, \Sigma, \zeta_{\alpha}^*, y_{0,\alpha}^*)$.
It is easily verified that ${\bf LOC}_{\alpha}^*$ constructed above is a safety local controller for $\alpha$,
i.e. condition (\ref{eq:loc_*}) holds.

\subsection{Main Result} \label{Subsec:MainResult}

By the same procedure as above, we construct for each controllable event $\alpha \in \Sigma_c$
a safety local controller ${\bf LOC}_\alpha^*$, and two liveness local controllers
${\bf LOC}_{\alpha,n}^\omega$ ($n = 1, 2$).
We shall verify that these local controllers collectively achieve the same controlled
behaviors as ${\bf SUP}^*$ in (\ref{eq:monosup*}) and ${\bf SUP}^\omega$ in (\ref{eq:monosup_w}).

\begin{thm} \label{thm:equ}
The set of safety local controllers $\{{\bf LOC}_\alpha^*\ |\ \alpha \in \Sigma_c\}$,
the set of liveness local controllers $\{{\bf LOC}_{\alpha,n}^\omega\ |\ \alpha \in \Sigma_c, n = 1,2\}$
constructed above solve the Supervisor Localization Problem for DES with infinite behavior, i.e.
\begin{align}
   L({\bf G}) &\cap\Big(\mathop \bigcap\limits_{\alpha \in \Sigma_{c}} L({\bf LOC}_{\alpha}^*) \Big) \notag \\
              &\cap \Big(\mathop \bigcap\limits_{\alpha \in \Sigma_{c}, n = 1, 2} L({\bf LOC}_{\alpha,n}^\omega) \Big)
               =L({\bf G}^{f^* \wedge f^\omega}) \label{eq:sub1_thm} \\
   S({\bf G}) &\cap \Big(\mathop \bigcap\limits_{\alpha \in \Sigma_{c}} lim(L({\bf LOC}_{\alpha}^*)) \Big) \notag \\
              &\cap \Big(\mathop \bigcap\limits_{\alpha \in \Sigma_{c}, n = 1,2} lim(L({\bf LOC}_{\alpha,n}^\omega)) \Big)
               = S({\bf G}^{f^* \wedge f^\omega})   \label{eq:sub2_thm}
\end{align}
where $L({\bf G}^{f^*\wedge f^\omega})$ and $S({\bf G}^{f^*\wedge f^\omega})$ respectively represent
the finite and infinite controlled behaviors of $\bf G$ under the control of ${\bf SUP}^*$ and ${\bf SUP}^\omega$
(as in (\ref{eq:sub1:implement}) and (\ref{eq:sub2:implement})).
\end{thm}

Theorem~\ref{thm:equ} confirms the control equivalence of the constructed local controllers
and supervisors ${\bf SUP}^*$ and ${\bf SUP}^\omega$. Indeed, according to the definition
of (safety and liveness) local controllers, the safety local controller ${\bf LOC}_\alpha^*$
enables/disables event $\alpha$ consistently with ${\bf SUP}^*$ and the liveness local
controllers ${\bf LOC}_{\alpha}^\omega$ enable/disable $\alpha$ consistently with ${\bf SUP}^*$.
Hence, to prove Theorem~\ref{thm:equ}, we show (i) the control equivalence of $\{{\bf LOC}_\alpha^*\ |\ \alpha \in \Sigma_c\}$
with ${\bf SUP}^*$ and (ii) the control equivalence of $\{{\bf LOC}_{\alpha,n}^\omega\ |\ \alpha \in \Sigma_c, n = 1,2\}$
with ${\bf SUP}^\omega$. The proof of the first part is similar to
that of the control equivalence of local controllers with the corresponding monolithic supervisor in \cite{CaiWon10a}.
The proof of the second part is particular, because at each local controller ${\bf LOC}_{\alpha,n}^\omega$,
we consider the disablement of $\alpha$ on only the strings $s \in C_n$. In the following, we provide the complete proof
of Theorem~\ref{thm:equ}.

%Equation (\ref{eq:sub1_equiv})
%which expresses the control equivalence of finite behaviors, is guaranteed by the
%localization procedure based on the concepts of control consistency and control cover
%defined in Subsection~\ref{Subsec:SupLoc}.
%Equation (\ref{eq:sub2_equiv}) states the control equivalence of infinite behaviors.
%Inspecting the above localization procedure,

%In the following, we presents the proof of Theorem~\ref{thm:equ}.

\vspace{1em}

\noindent {\it Proof of Theorem~\ref{thm:equ}}: (i) We prove the control equivalence of $\{{\bf LOC}_\alpha^*\ |\ \alpha \in \Sigma_c\}$
with ${\bf SUP}^*$, i.e.
\begin{align}
L({\bf G}) &\cap\Big(\mathop \bigcap\limits_{\alpha \in \Sigma_{c}} L({\bf LOC}_{\alpha}^*) \Big) = L({\bf G}^{f^*}) \label{eq:sub1:equiv_*}\\
S({\bf G}) &\cap \Big(\mathop \bigcap\limits_{\alpha \in \Sigma_{c}} lim(L({\bf LOC}_{\alpha}^*)) \Big) = S({\bf G}^{f^*}) \label{eq:sub2:equiv_*}
\end{align}
where $L({\bf G}^{f^*})$ and $S({\bf G}^{f^*})$ respectively represent the finite and
infinite controlled behavior of $\bf G$ under the control of ${\bf SUP}^*$.
The proof of (\ref{eq:sub1:equiv_*}) is similar to that of the control equivalence of
local controllers with the corresponding monolithic supervisor; for a detailed proof, see
Proposition 1 in \cite{CaiWon10a}.

With (\ref{eq:sub1:equiv_*}), equation (\ref{eq:sub2:equiv_*}) is immediate:
\begin{align*}
S({\bf G}^{f^*}) &= S({\bf G}) \cap lim(L({\bf G}^{f^*})) ~~ (\text{by (\ref{eq:infcontrolbehaiv})})\\
  &= S({\bf G}) \cap lim\Big(L({\bf G}) \cap \big(\mathop \bigcap\limits_{\alpha \in \Sigma_{c}}
   L({\bf LOC}_{\alpha}^*) \big)\Big)  ~~ (\text{by (\ref{eq:sub1:equiv_*})}) \\
 &= S({\bf G}) \cap lim(L({\bf G})) \cap \Big(\mathop \bigcap\limits_{\alpha \in \Sigma_{c}}
   lim(L({\bf LOC}_{\alpha}^*)) \Big)  ~~ \\
   &~~~~(\text{by Lemma~\ref{lem:equ_lim}}) \\
   &= S({\bf G}) \cap \Big(\mathop \bigcap\limits_{\alpha \in \Sigma_{c}}
   lim(L({\bf LOC}_{\alpha}^*)) \Big)  ~~  \\
   &~~~~(\text{because $S({\bf G}) \subseteq lim(L({\bf G}))$})
\end{align*}

(ii) We prove the control equivalence of $\{{\bf LOC}_{\alpha,n}^\omega\ |\ \alpha \in \Sigma_c, n = 1,2\}$
with ${\bf SUP}^\omega$, i.e.
\begin{align}
L({\bf G}^{f^*}) &\cap\Big(\mathop \bigcap\limits_{\alpha \in \Sigma_{c}, n = 1, 2} L({\bf LOC}_{\alpha,n}^\omega) \Big) = L({\bf G}^{f^*\wedge f^\omega}) \label{eq:sub1:equiv_w}\\
S({\bf G}^{f^*}) &\cap \Big(\mathop \bigcap\limits_{\alpha \in \Sigma_{c}, n = 1,2} lim(L({\bf LOC}_{\alpha,n}^\omega))= S({\bf G}^{f^* \wedge f^\omega}) \label{eq:sub2:equiv_w}
\end{align}
where $L({\bf G}^{f^*\wedge f^\omega})$ and $S({\bf G}^{f^*\wedge f^\omega})$ respectively represent the finite and
infinite controlled behavior of ${\bf G}^{f^*}$ under the control of ${\bf SUP}^\omega$.
According to (i), we only need to prove (\ref{eq:sub1:equiv_w}): equation (\ref{eq:sub2:equiv_w})
will be obtained from (\ref{eq:sub1:equiv_w}) and Lemma~\ref{lem:equ_lim}.
Since $L({\bf SUP}^\omega) = L({\bf G}^{f^*\wedge f^\omega})$ (according to (\ref{eq:monosup_w})),
we must prove $L({\bf G}^{f^*}) \cap(\mathop \bigcap\limits_{\alpha \in \Sigma_{c}, n = 1, 2}
L({\bf LOC}_{\alpha,n}^\omega) ) = L({\bf SUP}^\omega)$.

First, we show $L({\bf SUP}^\omega) \subseteq L({\bf G}) \cap (\mathop \cap\limits_{\alpha \in \Sigma_{c}, n = 1,2}
L({\bf LOC}_{\alpha,n}^\omega))$.
It suffices to show for all $\alpha \in \Sigma_c$ and $n = 1, 2$, $L({\bf SUP}^*)\subseteq
L({\bf LOC}_{\alpha,n}^\omega)$. Let $\alpha \in \Sigma_c$ and $s \in L({\bf SUP}^\omega)$;
we must show $s \in L({\bf LOC}_{\alpha,n}^\omega)$.
Write $s = \sigma_0,...,\sigma_m$; then $s \in L({\bf SUP}^\omega)$ and thus
there exist $x_0^\omega, ..., x_m^\omega \in X^\omega$ such that
\[\xi^\omega(x_j^\omega,\sigma_j) = x_{j+1}^\omega, j = 0, ..., m-1.\]
Then by the definition of $\mathcal{C}_{\alpha,n}^\omega$ and $\zeta_{\alpha,n}$,
for each $j = 0, ..., m - 1$, there exist
$i_j, i_{j+1} \in I$ such that
\[x_j^\omega \in X_{i_j}^\omega ~\&~ x_{j+1}^\omega \in X_{i_{j+1}}^\omega ~\&~ \zeta_{\alpha,n}(i_j,\sigma_j)=i_{j+1}.\]
So $\zeta_{\alpha,n}(i_0,\sigma_0...\sigma_n)!$, i.e. $\zeta_{\alpha,n}(i_0, s)!$.
Hence we have $s \in L({\bf LOC}_{\alpha,n})$.

Next, we prove $L({\bf G}) \cap (\mathop \cap\limits_{\alpha \in \Sigma_{c}, n = 1,2}
L({\bf LOC}_{\alpha,n}^\omega)) \subseteq L({\bf SUP}^\omega)$, by induction
on the length of strings.

For the {\it base case}, as it was assumed that $S({\bf G}^{f^*\wedge f^\omega})$ is
nonempty, it follows that the languages $L({\bf G}^{f^*})$, $L({\bf LOC}_{\alpha,n}^\omega)$
and $L({\bf SUP}^\omega)$ are all nonempty, the
empty string $\epsilon$ belongs to each.

For the {\it inductive step}, suppose that $s \in L({\bf G}^{f^*}) \cap (\mathop \cap\limits_{\alpha \in \Sigma_{c}, n = 1,2}
L({\bf LOC}_{\alpha,n}^\omega))$ implies $s \in L({\bf SUP}^\omega)$, and $s\sigma \in L({\bf G}^{f^*})
\cap (\mathop \cap\limits_{\alpha \in \Sigma_{c}, n = 1,2}
L({\bf LOC}_{\alpha,n}^\omega))$ for an arbitrary event $\sigma \in \Sigma$; we
must show that $s\sigma \in L({\bf SUP}^\omega)$. If $\sigma \in \Sigma_u$,
then $s\sigma \in L({\bf SUP}^\omega)$ because $L({\bf SUP}^\omega)$ is
$*$-controllable (by its $\omega$-controllability).

Otherwise, we have $\sigma \in \Sigma_c$ and there
exists a local controller ${\bf LOC}_{\alpha,n}^\omega$ for
$\sigma$: $\alpha = \sigma$; $n = 1$ if $s \in C_1 = pre(A)$, otherwise $n = 2$.
It follows from $s\alpha \in (\mathop \cap\limits_{\alpha \in \Sigma_{c}, n = 1,2}
   L({\bf LOC}_{\alpha,n}^\omega))$ that $s\alpha
\in L({\bf LOC}_{\alpha,n}^\omega)$ and $s \in L({\bf LOC}_{\alpha,n}^\omega)$. Namely,
$\zeta_{\alpha,n}^\omega(y_{0,\alpha,n}^\omega,s\alpha)!$ and $\zeta_{\alpha,n}^\omega(y_{0,\alpha,n}^\omega,s)!$.
Let $i := \zeta_{\alpha,n}^\omega(y_{0,\alpha,n}^\omega,s)$; then there exists
$j = \zeta_{\alpha,n}^\omega(i,\alpha)$. By the definition of $\zeta_{\alpha,n}^\omega$,
there exists $x^\omega,x'^\omega \in X_i^\omega$ and $x''^\omega \in X_j^\omega$ such that
$\xi^\omega(x_0^\omega,s) = x^\omega$ and $\xi^\omega(x'^\omega,\alpha) = x''^\omega$.
Since $x^\omega$ and $x'^\omega$ belong to the same cell
$X_i^\omega$, by the definition of control
cover they must be control consistent, i.e. $(x^\omega,x'^\omega)\in
\mathcal{R}_{\alpha,n}^\omega$. Thus $E_\alpha^\omega(x^\omega)\cdot D_{\alpha,n}^\omega(x'^\omega)=0$, which
implies $D_{\alpha,n}^\omega(x'^\omega) = 0$. The latter means that: either (a) $\xi^\omega(x^\omega,\alpha)!$ or (b) for all $t \in
C_n$ with $\xi^\omega(x_0^\omega,t) = x^\omega$, $\xi^*(x_0^*,t\alpha)$ is not
defined. Note that (b) is impossible because by hypothesis that $t \in L({\bf SUP}^\omega)$
and $t\alpha \in L({\bf G}^{f^*})$ we have $\xi^\omega(x_0^\omega,t)!$ and $\xi^*(x_0^*,t\alpha)!$.
Thus by (a), $\xi^\omega(\xi^\omega(x_0^\omega,s),\alpha)!$, and therefore $s\alpha \in L({\bf SUP}^\omega)$.

We have shown equations (\ref{eq:sub1:equiv_*}) and (\ref{eq:sub2:equiv_*}),
and equations (\ref{eq:sub1:equiv_w}) and (\ref{eq:sub2:equiv_w}). Combining them together,
we conclude that the equations (\ref{eq:sub1_thm}) and (\ref{eq:sub2_thm}) hold.
\hfill $\square$

From the proof of Theorem~\ref{thm:equ}, we see that the equivalences of infinite behaviors
(equations (\ref{eq:sub2:equiv_*}) and (\ref{eq:sub2:equiv_w})) are immediately derived from
their corresponding equivalences of finite behaviors (equations (\ref{eq:sub1:equiv_*}) and (\ref{eq:sub1:equiv_w}))
and Lemma~\ref{lem:equ_lim}. This confirms that the definitions of control consistency and control cover
need not contain any consistency relationship on infinite behavior. Thus the localization algorithm
(see \cite{CaiWon10a}) for DES with finite behavior can be easily adapted to construct local controllers in
Theorem~\ref{thm:equ} with suitable modifications: (i) using the current definition of control consistency and
control cover; (ii) for the localization of ${\bf SUP}^\omega$, we need to judge if a state $x$ in ${\bf SUP}^\omega$
can be arrived by a string $s \in C_n$ ($n = 1,2$). Assume that a $*$-automaton ${\bf C}_n = (Z,\Sigma,\eta,z_0)$ represents the $*$-language
$C_n$; then the above judgement can be realized by checking if state $x$ is in one of the state pairs of the product
of ${\bf SUP}^\omega$ and ${\bf C}_n$. The complexity of this step is $O(|X^\omega|\times |Z|)$.
We have known that the complexities of the localization algorithms for localizing ${\bf SUP}^*$
and ${\bf SUP}^\omega$ are $O(|X^*|^4)$ and $O(|X^\omega|^4)$ respectively, and thus the overall
complexity of the new localization procedure for DES with infinite behavior is $O(|X^*|^4 + |X^\omega|^4+|X^\omega|\times |Z|)$.
The Small Factory example in the next section will demonstrate the above result.

%% the end %%

\section{Case Study: Small Factory} \label{Sec:CaseStudy}

\subsection{Model Descriptions: plant and specifications}

\begin{figure}[!t]
\centering
    \includegraphics[scale = 0.9]{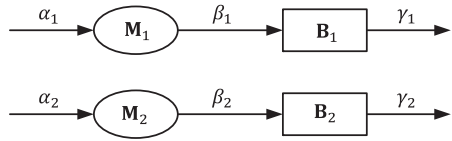}
\caption{Layout of Small Factory} \label{fig:SF_Layout}
\end{figure}

We illustrate the above supervisor localization for DES with infinite behavior by studying
a Small Factory example, taken from \cite[Chapt. 3]{Thistl91}. As displayed in
Fig.~\ref{fig:SF_Layout}, the plant to be controlled, denoted by $\bf SF$,
consists of two machines ${\bf M}_i$ ($i = 1, 2$) that are coupled with two
buffers ${\bf B}_i$ ($i = 1, 2$). The alphabet of event symbols for
$\bf SF$ is \[\Sigma = \{\alpha_1, \alpha_2,\beta_1,\beta_2,\gamma_1,\gamma_2\}.\]

\begin{figure}[!t]
\centering
    \includegraphics[scale = 0.8]{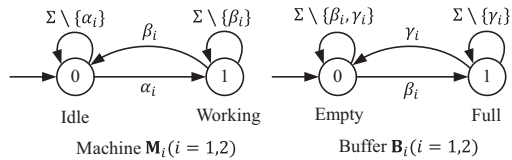}
\caption{$*$-automata representing finite behaviors of machines ${\bf M}_i$,
and buffers ${\bf B}_i$ ($i = 1, 2$). }
\label{fig:plant}
\end{figure}

The finite behavior of the plant is described as follows.
There are two routines in the plant. At each routine $i$ ($i = 1, 2$),
the machine ${\bf M}_i$ processes workpieces one at a time.
When ${\bf M}_i$ begins a job it acquires a workpiece from elsewhere in the factory
(event $\alpha_i$). Upon completing the ${\bf M}_i$ pushes the workpiece onto
buffer ${\bf B}_i$ (event $\beta_i$). Machines not shown in
Fig.~\ref{fig:SF_Layout} remove workpieces from buffer ${\bf B}_i$ for further
processing (event $\gamma_i$); we assume that some control mechanism
prevents such events from causing buffer ${\bf B}_i$ to ``underflow" -
supposing for the sake of simplicity that each buffer has only one slot.
The two machines and two buffers are modelled by the $*$-automata in
Fig. \ref{fig:plant}.

% The resultant strict alternation of $\alpha_i$ and $\beta_i$ is %represented
%by the $*$-automaton of Fig.~\ref{fig:Machine}.

\begin{figure}[!t]
\centering
    \includegraphics[scale = 0.8]{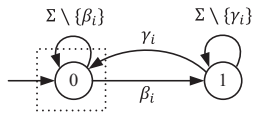}
\caption{B\"uchi automata ${\bf F}_i$ ($i = 1, 2$) representing the liveness assumptions
on each routine that every $\beta_i$ is eventually followed by an occurrence of $\gamma_i$.}
\label{fig:LiveAssumpt}
\end{figure}

The infinite behavior of the plant describes that removing workpieces from
the buffer are in continual operation, so that every occurrence of $\beta_i$
is {\it eventually} followed by an occurrence of $\gamma_i$. This behavior is
captured by the B\"uchi automata ${\bf F}_i$ ($i = 1,2$) of
Fig.~\ref{fig:LiveAssumpt}.

Now we have a complete model of the uncontrolled DES plant $\bf SF$:
the finite behavior is the intersection of the languages accepted by
the four $*$-automata in Fig. \ref{fig:plant},
i.e. \[L({\bf SF}) = L({\bf M}_1) \cap L({\bf M}_2) \cap L({\bf B}_1) \cap L({\bf B}_2);\]
the infinite behavior is the intersection of $lim(L({\bf SF}))$ with
the $\omega$-languages accepted by the two B\"uchi $*$-automata in
Fig. \ref{fig:LiveAssumpt},
i.e. \[S({\bf SF}) = lim(L({\bf SF})) \cap S({\bf F}_1) \cap S({\bf F}_2).\]

The plant under control must satisfy a number of specifications.
\begin{enumerate}[(S1)]
\item \label{spec:overflow} It should prevent buffer overflows: two occurrences
of $\beta_i$ should be separated by an occurrence of $\gamma_i$.
%The specifications are represented by $*$-automata ${\bf SPEC1}_i$
%($i = 1, 2$) as displayed in Fig.~\ref{fig:SafeSpec}.

\item \label{spec:mutual_exclusion} Because ${\bf M}_i$ ($i = 1,2$) employ the same
resources, they must not be allowed to operate simultaneously: $\alpha_i$
should not occur between successive occurrence of $\alpha_j$ and $\beta_j$.
%This specification is represented by the $*$-automaton ${\bf SPEC2}$ as
%displayed in Fig.~\ref{fig:SafeSpec}.

\item \label{spec:fairness} Because the ``mutual exclusion" requirement
(S\ref{spec:mutual_exclusion}) raises the possibility that one machine
may continually preempt the other, we add a liveness specification that
each machine operates {\it infinitely often}: in other words, each $\alpha_i$
should occur infinitely often.
%The specification is represented by the deterministic B\"uchi automaton
%displayed in Fig.~\ref{fig:MaxLiveSpec}.

\item \label{spec:minimal} The two routines in Fig.~\ref{fig:SF_Layout} {\it always}
work alternately, i.e. ${\bf M}_1$ (resp. ${\bf M}_2$) should not start (or restart) to work until
the workpiece in ${\bf B}_2$ (resp. ${\bf B}_1$) has been taken away. Here we assume that
initially ${\bf M}_1$ starts to work before ${\bf M}_2$. % and the specification
%is represented by the deterministic B\"uchi automaton displayed in Fig.~\ref{fig:MinLiveSpec}.

\end{enumerate}

\begin{figure}[!t]
\centering
    \includegraphics[scale = 0.8]{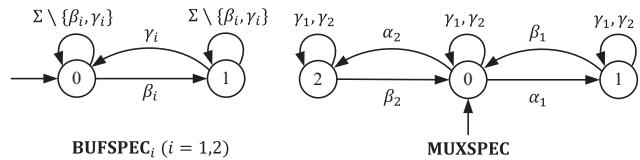}
\caption{Safety specifications: prevention of buffers' overflow represented by $*$-automata ${\bf BUFSPEC}_i$
($i = 1, 2$) and mutual exclusion requirement represented by $*$-automata $\bf MUXSPEC$}
\label{fig:SafeSpec}
\end{figure}

\begin{figure}[!t]
\centering
    \includegraphics[scale = 0.8]{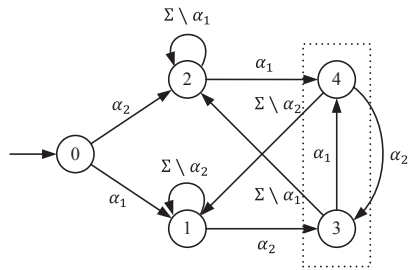}
\caption{Maximal legal liveness specification represented by B\"uchi automaton
${\bf MAXSPEC}$}
\label{fig:MaxLiveSpec}
\end{figure}

\begin{figure}[!t]
\centering
    \includegraphics[scale = 0.8]{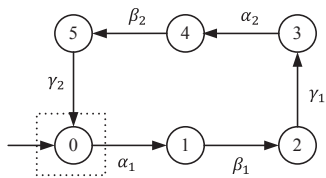}
\caption{Minimal acceptable liveness specification represented by
B\"uchi automaton ${\bf MINSPEC}$}
\label{fig:MinLiveSpec}
\end{figure}

Specifications (S\ref{spec:overflow}) and (S\ref{spec:mutual_exclusion}) are
represented by the $*$-automata ${\bf BUFSPEC}_i$ ($i = 1,2$) and $\bf MUXSPEC$ in Fig.~\ref{fig:SafeSpec}.
They describe finite behavioral requirements on the system, and thus are considered as safety specifications.
Let $E_s$ denote the overall safety specification,
i.e. \[E_s = L({\bf BUFSPEC}_1) \cap L({\bf BUFSPEC}_2) \cap L({\bf MUXSPEC}).\]
(S\ref{spec:fairness}) represented by the deterministic B\"uchi automaton ${\bf MAXSPEC}$
in Fig.~\ref{fig:MaxLiveSpec}, is considered as the maximal legal liveness specification,
i.e.
\[E_l = S({\bf MAXSPEC}).\]
(S\ref{spec:minimal}) represented by the deterministic B\"uchi automaton $\bf MINSPEC$,
is selected as the minimal acceptable liveness specification, i.e.
\[A = S({\bf MINSPEC}).\]

% the end

\subsection{Safety and Liveness Supervisors Synthesis}

There are two types of specifications imposed on the system $\bf SF$:
safety specification $E_s$ and liveness specifications $E_l$ and $A$.

\begin{figure}[!t]
\centering
    \includegraphics[scale = 0.8]{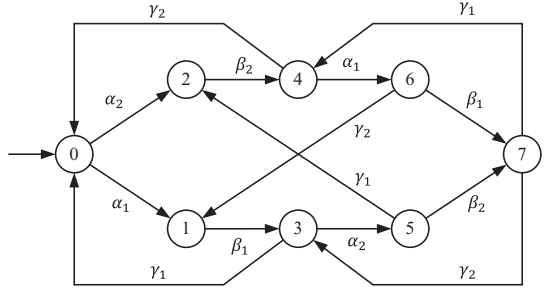}
\caption{Transition structure of ${\bf SF}^{f^*}$ and ${\bf SUP}^*$}
\label{fig:SUP}
\end{figure}

For safety specification, we compute as in (\ref{eq:monosup*}) a
safety supervisor ${\bf SUP}^* := (X^*, \Sigma, \xi^*, x_0^*)$ as displayed in Fig.~\ref{fig:SUP},
which has 8 states and 14 transitions.
The controlled behavior of ${\bf SF}$ under the control of ${\bf SUP}^*$
is represented by B\"uchi automaton ${\bf SF}^{f^*}$, i.e.
\begin{align*}
L({\bf SF}^{f^*}) &= L({\bf G}) \cap L({\bf SUP}^*)\\
S({\bf SF}^{f^*}) &= S({\bf SF}) \cap lim(L({\bf SUP}^*)).
\end{align*}
%The controlled behavior satisfies
%
${\bf SF}^{f^*}$ has the same transition structure with ${\bf SUP}^*$,
and the B\"uchi acceptance criterion accepting the language $S({\bf SF}^{f^*})$
is ${\mathcal{B}_{X^*}} = \{0,1,2,3,4\}$.
%The supervisor $f^*$ can be implemented by a $*$-automaton ${\bf SUP}^*$,
%i.e.
%\begin{align*}
%L({\bf SF}) \cap L({\bf SUP}^*) &= L({\bf SF}^{f^*}) \\
%S({\bf SF}) \cap lim(L({\bf SUP}^*)) &= S({\bf SF}^{f^*})
%\end{align*}
%${\bf SUP}^*$ has the same transition structure with ${\bf SF}^{f^*}$ as displayed in Fig.~\ref{fig:SUP}.

It is easily verified that the safety specifications (S\ref{spec:overflow}) and
(S\ref{spec:mutual_exclusion}) are both satisfied, i.e.
\[L({\bf SF}^{f^*}) = \sup\mathcal{C}^*(E_s\cap L({\bf SF})) \subseteq E_s.\]
However, there may exist
the case that one of machines, e.g. ${\bf M}_1$, may work recursively all the
time. In other words, ${\bf M}_1$ may preempt the start of ${\bf M}_2$ infinitely,
violating the liveness specification (S\ref{spec:fairness}).

\vspace{1em}

\begin{figure*}[!t]
\centering
    \includegraphics[scale = 0.8]{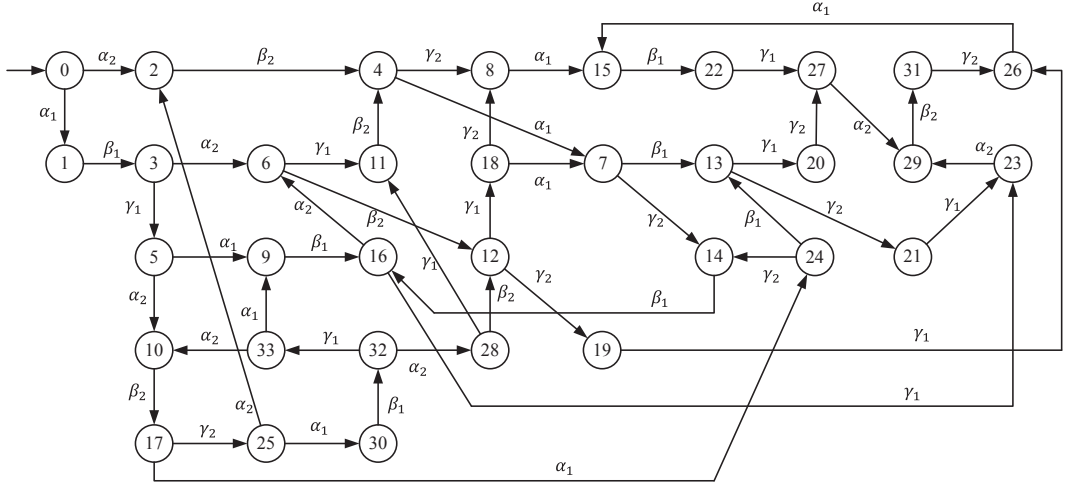}
\caption{Transition structure of ${\bf SF}^{f^*\wedge f^\omega}$ and ${\bf SUP}^\omega$}
\label{fig:SUPW}
\end{figure*}

For the maximal legal liveness specifications $E_l$ and minimal acceptable
liveness specification $A$, treating ${\bf SF}^{f^*}$ as the new plant to be controlled,
we construct as in (\ref{eq:monosup_w}) a liveness supervisor ${\bf SUP}^\omega$ 
as displayed in Fig.~\ref{fig:SUPW}, which has 34 states and 51 transitions.
The controlled behavior of ${\bf SF}^{f^*}$,
represented by B\"uchi automaton ${\bf SF}^{f^*\wedge f^\omega}$, i.e.
\begin{align*}
L({\bf SF}^{f^*\wedge f^\omega}) &= L({\bf SF}^{f^*}) \cap L({\bf SUP}^\omega)\\
S({\bf SF}^{f^*\wedge f^\omega}) &= S({\bf SF}^{f^*}) \cap lim(L({\bf SUP}^\omega)).
\end{align*}
${\bf SF}^{f^*\wedge f^\omega}$ has the same transition structure with ${\bf SUP}^\omega$, as displayed in Fig.~\ref{fig:SUPW},
and the B\"uchi acceptance criterion accepting the language $S({\bf SF}^{f^*\wedge f^\omega})$
is ${\mathcal{B}_{X^\omega}} = \{1,2,3,4,5,6,7,8,9,10,15,16,23,29,30\}$.
The readers are referred to Appendix~\ref{Append:Examp_SupSynth} for the detailed steps of
constructing ${\bf SUP}^\omega$. It is also verified that the controlled behavior
satisfies the given liveness specifications, i.e.
\begin{align*}
&A \subseteq S({\bf G}^{f^*\wedge f^\omega}) \subseteq E_l.
\end{align*}

Comparing the transition structure of ${\bf G}^{f^*}$ and ${\bf SUP}^\omega$,
we find that event $\alpha_1$ should be disabled at states
20, 23, 27, 31, and event $\alpha_2$ should be
disabled at states 8, 19, 22, 26.
To illustrate the control logic of supervisor ${\bf SUP}^\omega$, we consider the control actions
on event $\alpha_1$ at states 5 and 23. Since the plant of ${\bf SUP}^\omega$
is ${\bf SF}^{f^*}$, the finite controlled behavior must satisfy the safety specifications (S\ref{spec:overflow}) and
(S\ref{spec:fairness}), thus here we only consider the infinite behavior of the controlled plant.

First, $\alpha_1$ is enabled at state 5; the reason is as follows.
At state 5, only string $s := \alpha_1\beta_1\gamma_1$ has occurred,
namely, a workpiece has been taken by ${\bf M}_1$, deposited into ${\bf B}_1$ and
taken away from ${\bf B}_1$. At this stage, if $\alpha_1$ is enabled, there exists sublanguage
$L_{sub} = s\alpha_1\beta_1\gamma_1(\alpha_2\beta_2\gamma_2\alpha_1\beta_1\gamma_1)^\omega$
synthesized by ${\bf SUP}^\omega$, which satisfies the
liveness specification (S\ref{spec:fairness}).

However, the supervisor ${\bf SUP}^\omega$ chooses to disable event $\alpha_1$ at state 23; the reason is as follows.
Let $t = ss = \alpha_1\beta_1\gamma_1\alpha_1\beta_1\gamma_1$, and it is easily verified that
in ${\bf G}^{f^*}$, string $t$ re-visits state 0. As described in the above case, disabling event
$\alpha_1$ (on the contrary enabling event $\alpha_2$) may bring an infinite controlled behavior that satisfies the
liveness specification (S\ref{spec:fairness}). Hence, this disablement is correct.
Moreover, considering a general case when the string $s$ has occured $n < \infty$ times;
it is also safe for ${\bf M}_1$ to work again, because the supervisor can prevent ${\bf M}_1$
from starting to work, but permit ${\bf M}_2$ to start at $n+1$ times of occurrences of $s$.
However, we cannot enable event $\alpha_1$ infinitely, because
the infinite occurrences of string $s$ (i.e. $s^\omega$) will violate the liveness specification (S\ref{spec:fairness}).
Namely, event $\alpha_1$ must be disabled in a finite time; here ${\bf SUP}^\omega$ chooses to disable it
at string $t$. Hence, supervisor ${\bf SUP}^\omega$ is one, but not the unique supervisor for satisfying the liveness
specification (S\ref{spec:fairness}).

Now we have a safety supervisor ${\bf SUP}^*$ and a liveness supervisor ${\bf SUP}^\omega$,
whose finite and infinite controlled behaviors on the plant ${\bf SF}$ are represented by
$L({\bf SF}^{f^*\wedge f^\omega})$ and $S({\bf SF}^{f^*\wedge f^\omega})$, i.e.
\begin{align*}
L({\bf SF}^{f^*\wedge f^\omega}) &= L({\bf SF}) \cap L({\bf SUP}^*) \cap L({\bf SUP}^\omega)\\
S({\bf SF}^{f^*\wedge f^\omega}) &= S({\bf SF}^{f^*}) \cap lim(L({\bf SUP}^\omega)) \cap lim(L({\bf SUP}^\omega)).
\end{align*}
In the next subsection, we decompose the two supervisors into corresponding local controllers.
% the end

\subsection{Supervisor Localization}

\begin{figure}[!t]
\centering
    \includegraphics[scale = 0.95]{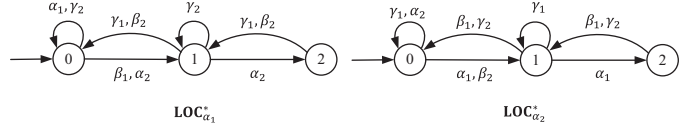}
\caption{Safety local controllers ${\bf LOC}_{\alpha_1}^*$ and ${\bf LOC}_{\alpha_2}^*$ for
controllable events $\alpha_1$ and $\alpha_2$ respectively}
\label{fig:LOC_*}
\end{figure}

There are two controllable events $\alpha_1$ and $\alpha_2$ in the plant ${\bf SF}$.
By applying the localization procedure in Section~\ref{Subsec:SupLocSafe},
we first get two safety local controllers ${\bf LOC}_{\alpha_1}^*$ and ${\bf LOC}_{\alpha_2}^*$
for controllable events $\alpha_1$ and $\alpha_2$ respectively, as shown in Fig.~\ref{fig:LOC_*}.

The control logic of ${\bf LOC}_{\alpha_1}^*$ is as follows. First, to prevent the overflow
of ${\bf B}_1$ (specification (S\ref{spec:overflow})), machine ${\bf M}_1$ is prohibited by ${\bf LOC}_{\alpha_1}^*$ to take a workpiece
from the source (i.e. event $\alpha_1$) when the buffer ${\bf B}_1$ is full, i.e. there exists a workpiece
in buffer ${\bf B}_1$, e.g. ${\bf LOC}_{\alpha_1}^*$ is at states 1 or 2.
Second, to satisfy the specification (S\ref{spec:mutual_exclusion}), event $\alpha_1$ should be disabled by
${\bf LOC}_{\alpha_1}^*$ between successive occurrences of $\alpha_2$ and $\beta_2$, e.g. ${\bf LOC}_{\alpha_1}^*$
is at states 1 and 2. Note that at state 1, the buffer may be empty and $\alpha_1$ is permitted to occur
without violating the specification (S\ref{spec:overflow}); however, at this state, $\alpha_1$ must be
disabled to prevent the violation of specification (S\ref{spec:mutual_exclusion}).

The control logic of ${\bf LOC}_{\alpha_2}^*$ is similar to that of ${\bf LOC}_{\alpha_1}^*$, but to disable
or enable event $\alpha_2$.

It is verified that ${\bf LOC}_{\alpha_1}^*$ and ${\bf LOC}_{\alpha_2}^*$ are control equivalent
to ${\bf SUP}^*$ in controlling the plant ${\bf SF}^*$, i.e.
\begin{align}
&L({\bf SF}) \cap L({\bf LOC}_{\alpha_1}^*) \cap L({\bf LOC}_{\alpha_2}^*) = L({\bf SF}^{f^*}) \label{eq:sub1:SF_equiv_*}\\
&S({\bf SF}) \cap lim(L({\bf LOC}_{\alpha_1}^*)) \cap lim(L({\bf LOC}_{\alpha_1}^*)) = S({\bf SF}^{f^*}). \label{eq:sub2:SF_equiv_*}
\end{align}

\begin{figure}[!t]
\centering
    \includegraphics[scale = 1]{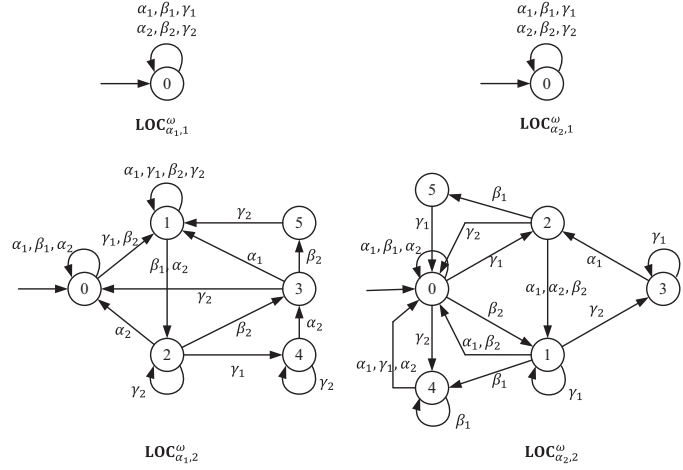}
\caption{Liveness local controllers ${\bf LOC}_{\alpha_1,1}^\omega$ and ${\bf LOC}_{\alpha_1,2}^\omega$ for
controllable event $\alpha_1$ and liveness local controllers ${\bf LOC}_{\alpha_1,1}^\omega$ and ${\bf LOC}_{\alpha_1,2}^\omega$ for $\alpha_2$}
\label{fig:LOC_w}
\end{figure}

Then, applying the localization procedure in Section~\ref{Subsec:SupLocLive}, we get two liveness local
controllers ${\bf LOC}_{\alpha_1,1}^\omega$ and ${\bf LOC}_{\alpha_1,2}^\omega$ for event $\alpha_1$,
and two liveness local controllers  ${\bf LOC}_{\alpha_2,1}^\omega$ and ${\bf LOC}_{\alpha_2,2}^\omega$ for event $\alpha_2$,
as displayed in Fig.~\ref{fig:LOC_w}.

Note that the liveness local controller ${\bf LOC}_{\alpha_1,1}^\omega$ (resp. ${\bf LOC}_{\alpha_2,1}^\omega$)
has only one state, namely event $\alpha_1$ need not be disabled at all the strings $s \in pre(A)$.
This control logic is consistent with ${\bf SUP}^\omega$: comparing the transition structures of ${\bf SUP}^*$
and ${\bf SUP}^\omega$, for all the states in ${\bf SUP}^\omega$ arrived by strings in $pre(A)$, event
$\alpha_1$ (resp. $\alpha_2$) is not disabled.

To illustrate the control logics of the liveness local controllers
${\bf LOC}_{\alpha_1,2}^\omega$ and ${\bf LOC}_{\alpha_2,2}^\omega$, we consider control action
of ${\bf LOC}_{\alpha_1,2}^\omega$ on $\alpha_1$ in the following cases. First, assume that the string
$s = \alpha_1\beta_1\gamma_1$ has occurred; ${\bf LOC}_{\alpha_1,2}^\omega$ arrives state 1, and
by inspecting the transition diagram of  ${\bf LOC}_{\alpha_1,2}^\omega$, $\alpha_1$ is enabled, consistent
with ${\bf SUP}^\omega$. Then, assume that the string $t = \alpha_1\beta_1\gamma_1\alpha_1\beta_1\gamma_1$
has occurred; now ${\bf LOC}_{\alpha_1,2}^\omega$ arrives state 4, and we can see that $\alpha_1$
is disabled by ${\bf LOC}_{\alpha_1,2}^\omega$. Again the control logic is consistent with that of
${\bf SUP}^\omega$.

It is also verified these four local controllers achieve the same controlled behavior with ${\bf SUP}^\omega$,
in controlling the plant ${\bf SF}^{f^*}$. i.e.
\begin{align}
&L({\bf SF}^{f^*}) \cap L({\bf LOC}_{\alpha_1,1}^\omega) \cap L({\bf LOC}_{\alpha_1,2}^\omega) \notag\\
&~~~~~~~~\cap L({\bf LOC}_{\alpha_2,1}^\omega) \cap L({\bf LOC}_{\alpha_2,2}^\omega) = L({\bf SF}^{f^*\wedge f^\omega}) \label{eq:sub1:SF_equiv_w} \\
&S({\bf SF}^{f^*}) \cap lim(L({\bf LOC}_{\alpha_1,1}^\omega)) \cap lim(L({\bf LOC}_{\alpha_1,2}^\omega)) \notag\\
&\cap lim(L({\bf LOC}_{\alpha_2,1}^\omega)) \cap lim(L({\bf LOC}_{\alpha_2,2}^\omega)) = S({\bf SF}^{f^*\wedge f^\omega}). \label{eq:sub2:SF_equiv_w}
\end{align}

Combining (\ref{eq:sub1:SF_equiv_*}) and (\ref{eq:sub1:SF_equiv_w}), (\ref{eq:sub2:SF_equiv_*}) and (\ref{eq:sub2:SF_equiv_w}),
we conclude that the above two safety local controllers ${\bf LOC}_{\alpha_1}^*$ and ${\bf LOC}_{\alpha_2}^*$
and the four liveness local controllers ${\bf LOC}_{\alpha_1,1}^\omega$, ${\bf LOC}_{\alpha_1,2}^\omega$,
${\bf LOC}_{\alpha_2,1}^\omega$ and ${\bf LOC}_{\alpha_2,2}^\omega$ achieve the same finite controlled behavior $L({\bf SF}^{f^*\wedge f^\omega})$
and infinite controlled behavior $S({\bf SF}^{f^*\wedge f^\omega})$, as ${\bf SUP}^*$ and ${\bf SUP}^\omega$, with respect
to the plant $\bf SF$.

\begin{figure}[!t]
\centering
    \includegraphics[scale = 0.8]{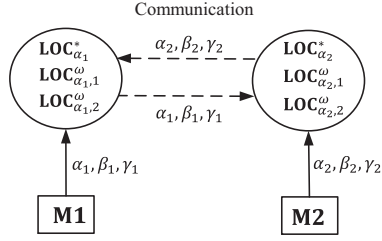}
\caption{Distributed control architecture for $\bf SF$}
\label{fig:DistriCtrlArch}
\end{figure}

Finally, with the derived local controllers, we build a distributed control architecture
for the small factory ${\bf SF}$; see Fig.~\ref{fig:DistriCtrlArch} of which the controlled behavior
satisfies the given specifications
(S\ref{spec:overflow}) - (S\ref{spec:minimal}).

% the end

\section{Conclusions} \label{Sec:Concl}

We have presented an extension of supervisor localization procedure to
solve the distributed control problem of multi-agent DES with infinite behavior.
We first employed Thistle and Wonham's supervisory control theory
for DES with infinite behavior to compute a safety supervisor (for safety specifications)
and a liveness supervisor (for liveness specifications), and implement
them by $*$-automata. Then we proposed a new supervisor localization theory to
decompose the safety and liveness supervisors into
a set of safety local controllers one for each controllable event, and a set of liveness local controllers
two for each controllable event, respectively.
Moreover, we have proved that the derived local controllers achieve
the same controlled behavior with the safety and liveness supervisors.
Finally, a Small Factory example has been presented for illustration.
In future research we shall consider the supervisory control and distributed control
of DES with infinite behavior under partial observation.

\appendices

\section{Effective Synthesis of Supervisor $f^\omega$} \label{Append:SupSynth}

To construct a complete and deadlock-free supervisor $f^\omega$ described in Section~\ref{Subsec:SupControl},
we need to compute $\sup\mathcal{C}^\omega(E_l)$ and $\inf\mathcal{F}^\omega(A)$
in advance. Without lose of generality, we assume that $A \subseteq E_l \subseteq S({\bf G})$.
If this assumption does not hold, we may replace $E_l$ and $A$
by $E_l' := E_l\cap S({\bf G})$ and $A' := E_l' \cap A$ respectively; $E_l'$ and
$A'$ will be treated as the new maximal legal specification and minimal acceptable
specification, but represent the same requirements on $\bf G$.

Define a deterministic Rabin-B\"{u}chi automaton
\begin{align} \label{eq:plant_Rabin}
\mathcal{A} = (Q', \Sigma, \delta', q_0', \{(R_p', I_p'): p \in P'\}, \mathcal{B}_{Q'})
\end{align}
such that the $*$-automaton $(Q', \Sigma, \delta', q_0')$ accepts the
$*$-behavior $L({\bf G}) \subseteq \Sigma^*$ of ${\bf G}$, the B\"{u}chi automaton
$(Q', \Sigma, \delta', q_0', \mathcal{B}_{Q'})$ accepts the $\omega$-behavior
$S({\bf G}) \subseteq \Sigma^*$ of ${\bf G}$, and the Rabin automaton
$(Q', \Sigma, \delta', q_0', \{(R_p', I_p'): p \in P'\})$ accepts the specification
$E_l \subseteq S({\bf G})$ (such an automaton can be constructed from the DES model ${\bf G}$ in (\ref{eq:plant}) and a Rabin automaton
accepting $E_l$). Note that if $S({\bf G})$ is $\omega$-closed, then by Proposition 5.6 in \cite{Thistl91}
it is redundant for the supervisor synthesis, and thus we can assume that
$S({\bf G}) = lim(L({\bf G}))$. In that case, it can be interpreted
as an absence of liveness assumptions in the modelling of the uncontrolled DES. Namely, in the DES
model ${\bf G}$ in (\ref{eq:plant}), we may drop the B\"{u}chi acceptance criterion.
Moreover, the computation of $\sup\mathcal{C}^\omega(E_l)$ is different from
that when the liveness assumptions are considered; for details, see \cite[Chapter 7]{Thistl91}.

First, the computation of $\sup\mathcal{C}^\omega(E_l)$ begins with computing the {\it controllability
subset} $C^\mathcal{A} \subseteq Q'$ of $\mathcal{A}$ in (\ref{eq:plant_Rabin}).
The subset $C^\mathcal{A}$, together with a map
\begin{align*}
\phi^{\mathcal{A}} : C^\mathcal{A} \rightarrow \Gamma,
\end{align*}
can be obtained by the subset construction algorithm in \cite{ThiWon94a},
which recursively applies the fixpoint calculus method \cite{EmeLei86}.
%Here $\Gamma := \{\gamma \subseteq \Sigma| \gamma \supseteq \Sigma_{u}\}$ is as defined as previous.
By Theorem 8.12 in \cite{Thistl91}, the deterministic Rabin automaton
\begin{align} \label{eq:ASUP}
\mathcal{A}_{sup} = (Q', \Sigma, \delta', q_0', \{(R_p', I_p'): p \in P'\}) ~| ~C^\mathcal{A}
\end{align}
accepts the $\omega$-language $\sup\mathcal{C}^\omega(E_l)$. Here the operator `$|$', restriction
to the subset $C^\mathcal{A} \subseteq Q'$, turns all other states into {\it degenerate}
%\footnote{A
%state $q \in Q$ is {\it degenerate} if there are transitions leaving $q$ but all of them lead
%back to $q$.}
states \cite{Thistl91} that do not satisfy the Rabin acceptance
condition $\{(R_p', I_p'): p \in P'\}$. Note that $\mathcal{A}_{sup}$ is a
deterministic Rabin automaton because $\mathcal{A}$ is deterministic and
the operator `$|$' does not change this property.

Second, to compute $\inf\mathcal{F}^\omega(A)$, we have by Proposition 5.8 in \cite{Thistl91},
$\inf\mathcal{F}^\omega(A) = clo(A) \cap S({\bf G})$.
Given a deterministic Rabin automaton %\footnote{Other $\omega$-automata including deterministic or nondeterministic
%Rabin, B\"uautomaton can be converted into a deterministic Rabin automaton}
which accepts the $\omega$-language $A$, we construct a deterministic
Rabin automaton $\mathcal{A}_{inf}$ accepting $\inf\mathcal{F}^\omega(A)$ by:
first construct an $\omega$-automaton accepting $clo(A)$, and then
intersect it with $\bf G$ which accepts the $\omega$-language $S({\bf G})$.

Now that we have
\begin{enumerate}[(i)]
\item a deterministic Rabin automaton $\mathcal{A}_{sup}$ accepting $\sup\mathcal{C}^\omega(E_l)$,
\item a controllability subset $C^{\mathcal{A}}$ together with a map $\phi^{\mathcal{A}} : C^\mathcal{A}
\rightarrow \Gamma$,
\item a deterministic Rabin automaton $\mathcal{A}_{inf}$ accepts
$\inf\mathcal{F}^\omega(A)$,
\end{enumerate}
we may check the existence of the supervisor $f^\omega$ and construct it if exists.

The existence verification of $f^\omega$ is equivalent to checking the containment
$\inf\mathcal{F}^\omega(A) \subseteq \sup\mathcal{C}^\omega(E_l)$;
it suffices to test the automaton $\mathcal{A}_{inf} \slash \mathcal{A}_{sup}$
accepting $\inf\mathcal{F}^\omega(A)\slash\sup\mathcal{C}^\omega(E_l)$
for emptiness. $\mathcal{A}_{inf} \slash \mathcal{A}_{sup}$ can be obtained
by intersect $\mathcal{A}_{inf}$ with the complement of $\mathcal{A}_{sup}$.
When the answer is yes, $f^\omega$ is constructed as follows.

\begin{enumerate}[(i)]

\item Write ${\mathcal{A}}_{sup} = (Q'', \Sigma, \delta'', q_0'', \{(R_p'', I_p''): p \in P''\})$;
${\mathcal{A}}_{sup}$ is the deterministic Rabin automaton accepting $E' := \sup\mathcal{C}^\omega(E_l)$.
Then a subset $Q_m'' \subseteq Q''$ can be computed such
that the $*$-automaton $(Q_m'', \Sigma, \delta'', q_0'')$ accepts $pre(E')$.
Because $E''$ is $*$-controllable with respect to $L({\bf G})$ (the finite behavior
represented by $\mathcal{A}$),
we may define the map $\phi_0: Q_m'' \rightarrow \Gamma$ as
\[\phi_0(q'') := \{\sigma \in \Sigma| \delta(q'',s\sigma) \in Q_m''\};\]
then by the proof of Proposition 4.4 \cite{Thistl91}, the map $f_0^\omega: \Sigma^* \rightarrow \Gamma$ given by
\[f_0^\omega(k) := \phi_0(\delta''(q_0'',k))\]
is a complete, deadlock-free supervisor for $\bf G$ that synthesizes the
$*$-language $pre(E')$ and the $\omega$-language $clo(E')\cap S({\bf G})$.

\item For each $k \in pre(E')$, let $q'' = \delta''(q_0'',k)$,
and let $\mathcal{A}_q'' = (Q'', \Sigma, \delta'', q'', \{(R_p'', I_p''): p \in P\})$.
Define $f_k^\omega: \Sigma^* \rightarrow \Gamma$ as
\[f_k^\omega(l/k) := \phi^{\mathcal{A}}(\delta''(q'', l/k)),\]
where $\phi^{\mathcal{A}}: Q'' \rightarrow \Gamma$ is obtained in the process of computing the
controllability subset $C^{\mathcal{A}}$ (according to (\ref{eq:ASUP}),
$Q''$ is isomorphic to $C^{\mathcal{A}}$). It is shown in Theorem 5.9 \cite{Thistl91} that
$f_k^\omega: \Sigma^* \rightarrow \Gamma$ is a complete, deadlock-free supervisor
for $\mathcal{A}_q''$, which synthesizes some $\omega$-sublanguage $E_k' \subseteq E'/k$.

\item Define the supervisor
\begin{align} \label{eq:sup_omega}
f^\omega:\Sigma^*\rightarrow \Gamma
\end{align}
according to:
\begin{eqnarray}
f^\omega(l) :=
\left\{
   \begin{array}{ll}
      f_0^\omega(l) & \text{if}~~ l \in pre(A) \notag\\
      f_k^\omega(l/k) & \text{if}~~ l \in k ~pre(E_k')~\text{where}~ k\in M \\
      \text{undefined} & \text{otherwise} \notag
   \end{array}
\right.
\end{eqnarray}
where $M$ is the set of all elements of $pre(E')\slash pre(\inf\mathcal{F}^\omega(A))$ of minimal length.
\end{enumerate}

It is shown by \cite[Theorem 5.3]{ThiWon94b} that $f^\omega:\Sigma^*\rightarrow \Gamma$ defined above
is a complete, deadlock-free supervisor for $\bf G$, and the controlled behaviors
of $\bf G$ satisfy conditions (\ref{eq:sub1:scpw}) and (\ref{eq:sub2:scpw}).

%% the end

\section{Supervisor Synthesis of ${\bf SUP}^\omega$ in Small Factory Example} \label{Append:Examp_SupSynth}

In the following, we adopt the supervisor synthesis procedure for
DES with infinite behavior in Appendix~\ref{Append:SupSynth} (reduced from
the synthesis procedure in \cite{Thistle1991}) to construct
a supervisor ${\bf SUP}^\omega$ satisfying the maximal legal liveness specifications $E_l$
and containing the minimal acceptable liveness specification $A$.
Recall that ${\bf SF}^{f^*}$ is the new plant to be controlled, with finite behavior
$L({\bf SF}^{f^*})$ and  infinite behavior $S({\bf SF}^{f^*})$.

\begin{figure}[!t]
\centering
    \includegraphics[scale = 0.72]{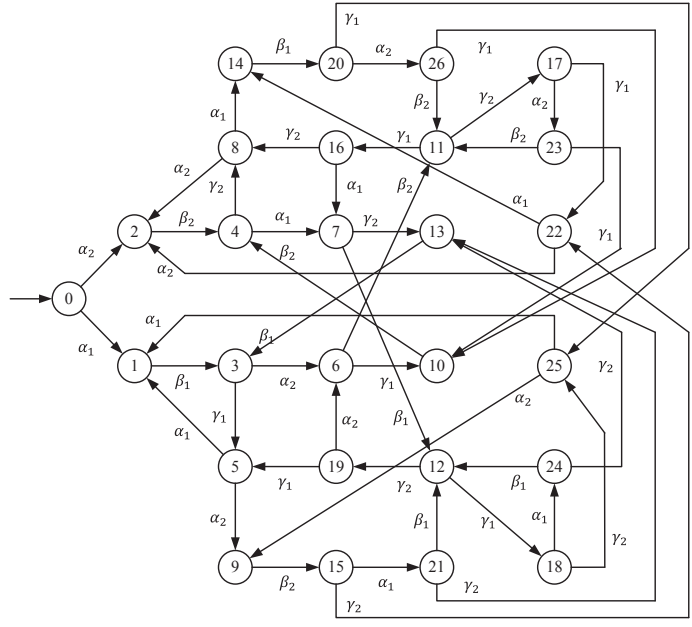}
\caption{Transition structure of Rabin-B\"uchi automaton $\mathcal{A}$}
\label{fig:RBA}
\end{figure}

{\bf Step} (i): Compute $\sup\mathcal{C}^\omega(E_l)$ and $\inf\mathcal{F}(A)$.
First, to compute $\sup\mathcal{C}^\omega(E_l)$, we construct a Rabin-B\"uchi
automaton
\[\mathcal{A} = (Q', \Sigma, \delta', q_0', \{(R_p', I_p'): p \in P'\}, \mathcal{B}_{Q'})\]
as in (\ref{eq:plant_Rabin}) such that the $*$-automaton $(Q', \Sigma, \delta', q_0')$ accepts the
$*$-behavior
\[L' := L({\bf SF}^{f^*}) \cap pre(E_l),\] the B\"{u}chi automaton
$(Q', \Sigma, \delta', q_0', \mathcal{B}_{Q'})$ accepts the $\omega$-behavior
\[S' := S({\bf SF}^{f^*}) \cap clo(E_l),\] and the Rabin automaton
$(Q', \Sigma, \delta', q_0', \{(R_p', I_p'): p \in P'\})$ accepts \[E' := S({\bf SF}^{f^*}) \cap E_l.\]
The transition structure of $\mathcal{A}$ is displayed in Fig.\ref{fig:RBA}, where
$Q' = \{0,1,...,26\}$, the B\"uchi acceptance criterion is $\mathcal{B}_{Q'} = \{1,2,3,4,5,6,7,8,9,14\}$, and
the Rabin acceptance criterion is $\{(R_1' = \{6,7,9,14\}, I_1' = Q')\}$.

It is easily verified that $E' \subseteq S' \subseteq lim(L')$ and thus by the controllability
subset construction algorithm proposed in \cite{ThiWon94a},
we compute the controllability subset $\mathcal{C}^{\mathcal{A}} = Q' = \{0,...,26\}$, together with
a map $\phi^{\mathcal{A}}: \mathcal{C}^{\mathcal{A}} \rightarrow \Gamma$, as listed
in Table~\ref{tab:statemap_phi}) (in the table, for each $q \in Q'$, $E_\delta(q) := \{\sigma \in \Sigma|\delta(q,\sigma)!\}$).

\begin{table*}
\footnotesize
\caption{State map  $\phi^{\mathcal{A}}: \mathcal{C}^{\mathcal{A}} \rightarrow \Gamma$} \label{tab:statemap_phi}
\begin{center}
\scalebox{1.0}{
\begin{tabular}{|c|c|c|c|c|c|c|c|c|c|c|c|c|c|c|}
\hline
$Q'$ & 0 & 1 &2 &3 &4 &5 &6 &7 & 8 & 9 & 10 & 11 & 12 & 13\\
\hline
$E_\delta$ & $\alpha_1,\alpha_2$ & $\beta_1$ & $\beta_2$ & $\gamma_1, \alpha_2$ & $\alpha_1, \gamma_2$ & $\alpha_1, \alpha_2$ & $\gamma_1, \beta_2$,& $\beta_1, \gamma_2$ &$\alpha_1,\alpha_2$ & $\beta_2$ & $\beta_2$ & $\gamma_1, \gamma_2$ & $\gamma_1, \gamma_2$ & $\beta_1$  \\
\hline
$\phi^{\mathcal{A}}$ & $\alpha_1,\alpha_2$ & $\beta_1$ & $\beta_2$ & $\gamma_1, \alpha_2$ & $\alpha_1, \gamma_2$ & $\alpha_2$ & $\gamma_1, \beta_2$,& $\beta_1, \gamma_2$ &$\alpha_1$ & $\beta_2$ & $\beta_2$ & $\gamma_1, \gamma_2$ & $\gamma_1, \gamma_2$ & $\beta_1$    \\
\hline
$Q'$ & 14 & 15 &16 &17 &18 &19 &20 &21 & 22 & 23 & 24 & 25 & 26 & \\
\hline
$E_\delta$ & $\beta_1$ & $\alpha_1,\gamma_2$ & $\alpha_1,\gamma_2$ & $\gamma_1, \alpha_2$ & $\alpha_1, \gamma_2$ & $\gamma_1, \alpha_2$ & $\gamma_1, \alpha_2$,& $\beta_1, \gamma_2$ &$\alpha_1,\alpha_2$ & $\gamma_1,\beta_2$ & $\beta_1,\gamma_2$ & $\alpha_1,\alpha_2$ & $\gamma_1, \beta_2$ &   \\
\hline
$\phi^{\mathcal{A}}$ & $\beta_1$ & $\gamma_2$ & $\alpha_1,\gamma_2$ & $\gamma_1$ & $ \gamma_2$ & $\gamma_1, \alpha_2$ & $\gamma_1$,& $\beta_1, \gamma_2$ &$\alpha_1$ & $\gamma_1,\beta_2$ & $\beta_1,\gamma_2$ & $\alpha_2$ & $\gamma_1, \beta_2$ &     \\
\hline

\end{tabular}
}
\end{center}
\end{table*}

Then, from ${\mathcal{A}}$ and its controllability subset $\mathcal{C}^{\mathcal{ A}}$,
we construct as in (\ref{eq:ASUP}) a Rabin automaton ${\mathcal{ A}}^\omega$ accepting
$\sup\mathcal{C}^\omega(E')$.  Since $\mathcal{C}^{\mathcal{ A}} = Q'$,
${\mathcal{ A}^\omega} := (Q', \Sigma, \delta', q_0', \{(R_p', I_p'): p \in P'\})$,
namely
\[\sup\mathcal{C}^\omega(E') = S({\mathcal{ A}}^\omega) = E'.\]
Hence $E'$ is $\omega$-controllable, but need not be $\omega$-closed; indeed, $E'$ is
not $\omega$-closed, because $s = \alpha_2\beta_2\alpha_1\beta_1\\(\gamma_1\alpha_2\beta_1)^*\gamma_2\gamma_1(\alpha_1\beta_1\gamma_1)^\omega$
belongs to $E'$, but $clo(s)$ does not.

Finally, inspecting the transition structure of $\bf MINSPEC$ representing the minimal
acceptable language $A$, we have $A = S({\bf MINSPEC}) = lim(L({\bf MINSPEC})) = clo(S({\bf MINSPEC})) = clo(A)$,
where $clo(A)$ is the $\omega$-closure of $A$ (for definition see (\ref{eq:def_clo}));
thus \[\inf\mathcal{F}^\omega(A) = clo(A) \cap S' = A.\]

It is verified that \[\inf\mathcal{F}^\omega(A) \subseteq \sup\mathcal{C}^\omega(E').\]
Hence by \cite[Theorem 5.3]{ThiWon94b} there exists a complete, deadlock-free
supervisor $f^\omega$ such that $A \subseteq S({\bf SF}^{f^*\wedge{f^\omega}}) \subseteq E' \subseteq E$,
where ${\bf SF}^{f^*\wedge{f^\omega}}$ is the new plant ${\bf SF}^{f^*}$ under the control of $f^\omega$.

{\bf Step} (ii): Synthesize supervisor $f^\omega:\Sigma^*\rightarrow \Gamma$. We first construct a supervisor
$f_0^\omega: \Sigma^*\rightarrow \Gamma$ according to $f_0^\omega(s) = E_\delta(\delta(q_0,s))$,
which synthesizes $*$-language $E'' := pre(\sup\mathcal{C}^\omega(E'))$.
Then, for each $k \in E''$, let $q = \delta'(q_0',k)$;
we construct a supervisor $f_k^\omega: \Sigma^* \rightarrow \Gamma$ according to
$f_k^\omega(l/k) = \phi^{\mathcal{A}}(\delta'(q,l/k))$, which synthesizes some sublanguage $E_k'' \subseteq E''/k$.
Next, write ${\bf MINSPEC} = (Z,\Sigma,\zeta,z_0,\mathcal{B}_Z)$ where $Z = \{0,1,...,5\}$, $\zeta$ is a partial function
as displayed in Fig.~\ref{fig:MinLiveSpec}, and $\mathcal{G}_Z = \{0\}$; we extend the transition function of
${\bf MINSPEC}$ to total function by adding an extra state 6,
(i.e. $Z = \{0,1,...,6\}$) and adding the transition $(z,\sigma,6)$ for every state $z \in Z$ including
$6$ if $\sigma$ is not defined at $z$.

Now, we are ready to construct a supervisor $f^\omega:\Sigma^*\rightarrow \Gamma$ (as in (\ref{eq:sup_omega})),
according to:
\begin{eqnarray}
f^\omega(l) :=
\left\{
   \begin{array}{ll}
      f_0^\omega(l) & \text{if}~~ l \in pre(A) \notag\\
      f_k^\omega(l/k) & \text{if}~~ l \in k ~pre(E_k'')~\text{where}~ k\in M \\
      \text{undefined} & \text{otherwise} \notag
   \end{array}
\right.
\end{eqnarray}
where $M$ is the set of all elements of $pre(E'')\slash pre(A)$ of minimal length.
The supervisor $f^\omega$ can be expressed by the state map $\psi:(Q',Z)\rightarrow \Gamma$
(as listed in Table~\ref{tab:supf}) in the form of $f^\omega(s) = \psi((\delta'\times\zeta((q_0',z_0),s))$.
In Table~\ref{tab:supf}, $E_{\delta',\zeta}(q,z) := \{\sigma\in\Sigma|\delta'(q,\sigma)! \& \zeta(z,\sigma)!\}$.
Note that if a string $l \in pre(A)$, then it arrives the state pairs $(q,z)$ with $z = 0,..., 5$
and in this case, $f^\omega(l) = f_0^\omega(l)$; otherwise, it arrives the state pairs
$(q,z)$ with $z = 6$ and in this case, $f^\omega(l) = f_k^\omega(l/k)$.
\begin{table*}
\footnotesize
\caption{State map  $\psi: Q' \times Z \rightarrow \Gamma$} \label{tab:supf}
\begin{center}
\scalebox{1.0}{
\begin{tabular}{|c|c|c|c|c|c|c|c|c|c|c|c|c|}
\hline
$(Q'\times Z)$ & (0,0) & (1,1) &(2,6) &(3,2) &(4,6) &(5,3) &(6,6) &(7,6) & (8,6) & (1,6) & (9,4) & (10,6)\\
\hline
$E_{\delta,\zeta}$ & $\alpha_1,\alpha_2$ & $\beta_1$ & $\beta_2$ & $\gamma_1, \alpha_2$ & $\alpha_1, \gamma_2$ & $\alpha_1, \alpha_2$ & $\gamma_1, \beta_2$,& $\beta_1, \gamma_2$ &$\alpha_1,\alpha_2$ & $\beta_1$ & $\beta_2$ & $\beta_2$   \\
\hline
$\psi$ & $\alpha_1,\alpha_2$ & $\beta_1$ & $\beta_2$ & $\gamma_1, \alpha_2$ & $\alpha_1, \gamma_2$ & $\alpha_1, \alpha_2$ & $\gamma_1, \beta_2$,& $\beta_1, \gamma_2$ &$\alpha_1$ & $\beta_1$ & $\beta_2$ & $\beta_2$   \\
\hline
$(Q'\times Z)$ & (11,6) & (12,6) &(13,6) &(14,6) &(3,6) &(15,5) &(16,6) &(17,6) & (18,6) & (19,6) & (20,6) & (5,6)\\
\hline
$E_\delta$ & $\gamma_1,\gamma_2$ & $\gamma_1,\gamma_2$ & $\beta_1$ & $\beta_1$ & $\gamma_1,\alpha_2$ & $\alpha_1,\gamma_2$ & $\alpha_1,\gamma_2$,& $\gamma_1,\alpha_2$ &$\alpha_1,\gamma_2$ & $\gamma_1,\alpha_2$ & $\gamma_1,\alpha_2$ & $\alpha_1,\alpha_2$  \\
\hline
$\psi$ & $\gamma_1,\gamma_2$ & $\gamma_1,\gamma_2$ & $\beta_1$ & $\beta_1$ & $\gamma_1,\alpha_2$ & $\alpha_1,\gamma_2$ & $\alpha_1,\gamma_2$,& $\gamma_1$ &$\gamma_2$ & $\gamma_1,\alpha_2$ & $\gamma_1$ & $\alpha_2$  \\
\hline
$(Q'\times Z)$ & (21,6) & (22,0) &(22,6) &(23,6) &(24,6) &(25,6) &(26,6) &(9,6) & (14,1) & (15,6) & (20,2) & (25,3)\\
\hline
$E_\delta$ & $\beta_1,\gamma_2$ & $\alpha_1,\alpha_2$ & $\alpha_1,\alpha_2$ & $\gamma_1, \beta_2$ & $\beta_1, \gamma_2$ & $\alpha_1, \alpha_2$ & $\gamma_1, \beta_2$,& $\beta_2$ &$\beta_1$ & $\alpha_1,\gamma_2$ & $\gamma_1,\alpha_2$ & $\alpha_1,\alpha_2$  \\
\hline
$\psi$ & $\beta_1,\gamma_2$ & $\alpha_1,\alpha_2$ & $\alpha_1$ & $\gamma_1, \beta_2$ & $\beta_1, \gamma_2$ & $\alpha_2$ & $\gamma_1, \beta_2$,& $\beta_2$ &$\beta_1$ & $\gamma_2$ & $\gamma_1,\alpha_2$ & $\alpha_1,\alpha_2$  \\
\hline

\end{tabular}
}
\end{center}
\end{table*}

Under the control of $f^\omega$, as described in Section~\ref{Sec:ProbForm},
the behavior of the new plant ${\bf SF}^{f^*}$ can be represented
by a deterministic B\"uchi automaton ${\bf SF}^{f^*\wedge f^\omega}$, as displayed in Fig.\ref{fig:SUPW}
(the B\"uchi acceptance criterion is $\mathcal{B} = \{1,2,3,4,5,6,7,8,9,10,15,16,23,29,30\}$).
It is easily verified that the controlled behavior satisfies all the
specifications in the following sense:
\begin{align*}
&L({\bf SF}^{f^*\wedge f^\omega})\subseteq E_s ~~~~~~~~ \mbox{(safety specifications (S\ref{spec:overflow}) and S\ref{spec:mutual_exclusion})} \\
&A \subseteq S({\bf SF}^{f^*\wedge f^\omega}) \subseteq E_l ~~ \mbox{(liveness specifications (S\ref{spec:fairness}) and (S\ref{spec:minimal}))}
\end{align*}

{\bf Step} (iii): Implement $f^\omega$ by $*$-automaton ${\bf SUP}^\omega$. The above
function-based supervisor $f^\omega$ can be implemented by a $*$-automaton ${\bf SUP}^\omega$
as displayed in Fig.~\ref{fig:SUPW}, i.e.
\begin{align*}
L({\bf SF}^{f^*}) \cap L({\bf SUP}^\omega) &= L({\bf SF}^{f^*\wedge f^\omega}) \\
S({\bf SF}^{f^*}) \cap lim(L({\bf SUP}^\omega)) &= S({\bf SF}^{f^*\wedge f^\omega})
\end{align*}
${\bf SUP}^\omega$ has the same transition structure with ${\bf SF}^{f^*\wedge f^\omega}$.

% the end

\bibliographystyle{IEEEtran}
\bibliography{SCDES_Ref,InfiniteBehav}

\end{document}